\documentclass[fleqn,usenatbib]{mnras}

\usepackage{newtxtext,newtxmath}
\usepackage[T1]{fontenc}
\usepackage{ae,aecompl}

\usepackage[pdftex]{graphicx}% Including figure files
\usepackage{amsmath}% Advanced maths commands
\usepackage{amssymb}% Extra maths symbols
\usepackage[center]{caption}
\usepackage[dvipsnames]{xcolor}
\usepackage{epstopdf}
\title[A method for quantifying satellite abundance]{A hierarchical clustering method for quantifying satellite abundance}

\author[C. Xi et al.]
{Chengyu Xi,$^{1, 2}$\thanks{E-mail: cxi@uwaterloo.ca}
James E. Taylor,$^{1, 2}$\thanks{E-mail: taylor@uwaterloo.ca}
\\
$^{1}$Department of Physics and Astronomy, University of Waterloo, 200 University Avenue West, Waterloo, Ontario, N2L 3G1, Canada\\
$^{2}$Waterloo Centre for Astrophysics, University of Waterloo,  200 University Avenue West, Waterloo, Ontario, N2L 3G1, Canada
}

\date{Accepted XXX. Received YYY; in original form ZZZ}

\pubyear{2020}

\begin{document}
\label{firstpage}
\pagerange{\pageref{firstpage}--\pageref{lastpage}}
\maketitle

\begin{abstract}
We present a new method for quantifying the abundance of satellites around field galaxies and in groups. The method is designed to work with samples, such as local photometric redshift catalogues, that do not have full spectroscopic coverage, but for which some redshift or distance information is available. It consists of identifying the galaxies most likely to be centrals, and using the clustering signal around them as a template to iteratively decompose the full population into satellite and central populations. In that sense it is similar to performing crowded-field photometry, after having first used isolated stars to determine the point spread function of the image. The method does not identify individual satellites or centrals conclusively, but assigns a probability to each galaxy of being one or the other. Averaged over a large sample, it provides a statistical estimate of satellite abundance, even in crowded fields with large redshift uncertainties. We test the method using data from the COSMOS field, which includes a large set of local objects with accurate photometric redshifts. We measure satellite abundance as a function of central stellar or halo mass, as well as the satellite luminosity function, and find results consistent with previous studies, but extending over a broader range of central masses. We also consider a number of possible systematic uncertainties in the method, and show that they are generally smaller than our random errors. Having presented the method in this paper, we will use it to study the properties of the satellite populations in a forthcoming one.
\end{abstract}


%%%% Section 1 %%%%

\section{Introduction} 
\label{sec:intro}
In the current picture of hierarchical structure formation, cold dark matter (CDM) haloes merge together to form progressively larger systems as the Universe evolves. Smaller dark matter halos often survive accretion onto larger systems, leaving a population of distinct `subhalos' within CDM halos. While the gas that cools and settles into the centre of the main halo will contribute to the growth of a central galaxy, smaller galaxies that formed within subhalos before they merged can survive as distinct satellite galaxies, subject to a broad range of environmental effects that may transform or destroy them, including tidal heating or stripping, encounters, and internal or external feedback. The observed abundance of satellites in the local Universe provides a detailed test of this complex picture and gives important insights into the overall effect of environment on galaxy formation.

The dominant galaxies of the Local Group (LG), the Milky Way (MW) and M31, have the best studied satellite populations in the Universe. Recent surveys have discovered many new, faint members of the LG \citep[e.g.][]{Bechtol2015, Drlica-Wagner2015, Koposov2015}, such that the total abundance of LG satellites can be estimated with increasing confidence \citep{Newton2018}. Over the past two decades, however, several points of tension have arisen between the observed population of LG satellites and that expected from theory. The most famous is the ``missing satellite problem'', which contrasts the small number of observed satellites with the large number of dark structures predicted by theory \citep{Klypin99, Moore09}. A second,  ``too-big-to-fail'' problem contrasts the low central densities estimated in the massive satellites of the MW with the much higher densities expected from theory \citep{Boylan-Kolchin2011-TBTF}. There may be other tensions as well, in the radial clustering \citep[e.g.][]{Kravtsov2004, Taylor2004} or 3-D spatial distribution \citep[e.g.][]{Pawlowski2015}. We refer the reader to \citet{Bullock2017} for a detailed review of these challenges. 

Many solutions have been proposed to resolve the tensions between theory and observations of the LG satellites, including internal feedback due to star formation \citep[e.g.][]{Dekel1986,Maschenko2008,Governato2010,Wetzel2016}, the effects of global \citep[e.g.][]{Bullock2000,Gnedin2006} and/or inhomogeneous \citep[e.g.][]{Lunnan2012} reionization, tidal or other environmental effects \citep[e.g.][]{Taylor2001,Mayer2006,Lokas2012}, or modifications to the underlying dark matter model such as warm dark matter \citep[e.g.][]{Maccio2010,Anderhalden2013,Kennedy2014,Lovell2014}, self-interacting dark matter \citep[e.g.][]{Spergel2000,Fry2015,Elbert2015}, or fuzzy dark matter 
\citep[e.g.][]{Nadler2019}.
 
 There remains, however, the important question of whether the MW and/or LG satellites are representative of all satellite populations. Many observational studies have show that the MW is {\it not} typical in having two LMC/SMC-like satellites \citep{GuoQuan11, LiuL2011, Strigari2012, Robotham2012, ST14}, and similar conclusions have been suggested by numerical simulations \citep{Boylan-Kolchin2010,Busha2011,KangX2016,ZhangD2019}. The ongoing Satellites Around Galactic Analogs (SAGA) survey \citep{Geha17} has also shown that there is a large variation in satellite populations from system to system. Theoretical models predict that the abundance of halo substructure should vary more than expected from Poisson statistics alone, and should be correlated with the formation redshift of the system \citep[e.g.][]{Jiang2017, Chua2017}. These complications caution us from relying too heavily on the properties of a single system to constrain models of galaxy formation. To determine whether the LG is representative, and to understand satellite properties across a broad range of environments, we should seek out satellites around as large a sample as possible of central galaxies.

Identifying satellites and distinguishing them from foreground or background systems requires some form of distance information. The main approaches in the literature include: (1) the use of existing complete spectroscopy to identify satellites around the nearest and brightest systems \citep[e.g.][]{Yang07}; (2) dedicated spectroscopic campaigns to obtain spectroscopy for fainter targets around a smaller number of selected systems \citep{Geha17}; (3) the use of photometric distance estimates from techniques such as the tip of the red giant branch (TRGB) \citep[e.g.][]{Carlin2016,Danieli2017, Cohen2018,Danieli2019}, or surface brightness fluctuations \citep[e.g.][]{vanDokkum2018, Carlsten2019}; (4) statistical abundance measurements based on clustering \citep[e.g.][]{LiuL2011, GuoQuan11,GuoQuan2012, Strigari2012, WangWT12, WangWT14, Sales2013, ST14, Xi18}. The four approaches have different strengths and weaknesses. Method (1) requires only existing data, but is restricted to the brightest satellites in the nearest systems, and may also suffer from incompleteness due to fibre positioning limitations in dense fields \citep[e.g.][]{GuoH2012,Smith2019}. Method (2) is extremely expensive in terms of observing time, and thus limited to small numbers of systems. Method (3) is restricted to very nearby systems [$<$20 Mpc], whose virial radii subtend large angles on the sky, making complete coverage difficult. Method (4) cannot confirm  individual galaxies as satellites or centrals; it has been very successful, however, in making measurements of the average satellite abundance, and is the least resource-intensive method of the four {\it a priori}.

Clustering-based methods have generally been applied to samples at redshifts $\sim$0.05--0.2, selected from the Sloan Digital Sky Survey (SDSS -- \citet{SDSS}; e.g.~\citet{LiuL2011, GuoQuan11,GuoQuan2012, Strigari2012, WangWT12, WangWT14, Sales2013}). A different strategy was adopted by \citet{ST14}, who focussed on very nearby systems (out to 42 Mpc). This allowed them to estimate the abundance of intrinsically faint satellites, at the expense of significant background contamination. They used a selection technique based on galaxy structural properties (mainly apparent size) to reduce the background contamination and boost the signal-to-noise ratio (SNR) of the clustering measurement. The technique was further developed and tested in \citet{Xi18}, using a broader range of morphological cuts. An optimized version was shown to be effective up to $z \sim 0.15$, far beyond the range considered in \citet{ST14}.

These previous clustering-based studies have generally considered samples of primaries that are clearly isolated, in the sense that they have no brighter companion within fixed projected and line-of-sight separations. This approach works well for bright, massive primaries, but becomes inefficient for less luminous ones. By dropping isolation cuts, \citet{Xi18} were able to detect a clear clustering signal and constrain satellite abundance using only observations from the fairly small COSMOS field, but this resulted in a broad selection of primaries, including many systems with overlapping virial regions. As a result, the interpretation of their results remains slightly unclear, relative to previous studies, as not all of their primaries are true central galaxies.

In this work, we introduce a new method to deal with the complications of overlapping systems and crowded fields. We start by identifying the subset of galaxies in a sample most likely to be true central galaxies, using a hierarchical search in which galaxies are checked for isolation in order of decreasing stellar mass, with isolation criteria that scale with the estimated virial radius of the system. The cross-correlation function of the sample with respect to this set of most likely primaries provides an initial template for the clustering signal. This template is used to estimate the probability that {\it any} member of the sample is a primary or a secondary. Finally, we can iterate through the last two steps, recalculating a probability-weighted cross-correlation function and the modified primary/secondary probabilities until convergence. The final primary/secondary probabilities for the whole sample then allow us to estimate satellite abundance, luminosity functions, and other distributions of secondary properties. Note that we have developed and optimized our method for low redshift samples. Some of our assumptions may need modification, in order to apply the method at higher redshifts.

In this paper we present the method and give some simple estimates of satellite abundance; in a forthcoming paper we will study the properties of the detected satellite populations in more detail. The paper is structured as follows. In Section \ref{sec:data} we describe our data selection, including the basic cuts that define our initial sample. In Section \ref{sec:clustering scale} we measure the clustering signal and use it to define a ``region of interest'' around each primary likely to contain most genuine satellites. In Section \ref{sec:p-s selection} we describe our iterative method for estimating primary and secondary probabilities for each galaxy. In Section \ref{sec:results} we present our main results on satellite abundance. In Section \ref{sec:test} we test the method for possible systematic uncertainties. Finally, in Section \ref{sec:conclusion} we summarize our results and discuss future prospects for this new method. 

%%%% Section 2 %%%%

\section{Data--COSMOS}
\label{sec:data}

The satellite galaxies we can hope to detect around a low-redshift primary (at most a few tens per system, based on abundances in the Local Group) will be seen in projection with a much larger number of foreground and background galaxies (on the order of thousands) that are not physically associated with the primary. Precise distance information is essential for separating true satellites from this foreground/background population. Spectroscopic redshifts are ideal for this purpose, but impractical for large samples. For instance, if we want to search for satellites brighter than $-18$ in absolute magnitude out to a redshift of 0.2, this requires distance information for galaxies down to an apparent magnitude of roughly 22. However, with a few exceptions \citep[e.g.][]{Geha17}, wide-field spectroscopic catalogues are usually only complete down to an apparent magnitude of 17 to 18, far from the depth required. Thus, using photometric redshifts (``photo-zs'') is the only realistic solution. The COSMOS field features high quality photo-zs generated from 30+ deep bands \citep{Scoville07,Ibert13_catalog, Laigle16}, making it an ideal place to test our method.

\subsection{The COSMOS photometric redshift catalogue}
\label{sec:COSMOS}

COSMOS is a deep ($AB \sim 25$--$26$), multi-wavelength (0.25 $\mu m$--24 $\mu m$) survey covering a 2 deg$^2$ equatorial field \citep{Scoville07}. The multi-wavelength imaging includes Hubble Space Telescope (HST) imaging with the Advanced Camera for Survey (ACS) and follow-up observations from many other facilities across a wide range of wavelengths -- X-ray, UV, optical/IR, FIR/submillimeter and radio \citep{Scoville07}. In this paper, we will use a recently updated photometric redshift catalogue \citep[][`COSMOS 2015' hereafter]{Laigle16} for our analysis. The main improvement of this catalogue compared to the previous releases is the addition of new, deeper NIR and IR data from the second data release (DR2) of the UltraVISTA and SPLASH \citep[\textit{Spitzer} Large Area Survey with Hyper-Suprime-Cam][]{Miyazaki12_HyperSupirmeCam} projects. Compared to the first data release (DR1) of UltraVISTA, the exposure time of DR2 was significantly longer \citep{McCracken12}, providing the deeper IR and NIR data as well as better SNRs \citep{Laigle16}. On the other hand, the DR 2 data only covers a part (namely `ultra-deep stripes', roughly 0.6 deg$^2$) of the COSMOS field. This causes a slight inconsistency in depth and SNR across the field, which we will address below by applying a magnitude cut. 

The COSMOS photo-zs were derived using $\chi^2$ template fitting, as described in \citet{Mobasher07} and \citet{Ilbert09}. The Spectral Energy Distribution (SED) templates used in the COSMOS 2015 catalogue include a set of 31 spiral and elliptical galaxies from \citet{Polletta07} and a set of templates for young blue star-forming galaxies generated using \citet{Bruzual03} models. Given the updated NIR and IR data and two additional star-forming galaxy templates, \citet{Laigle16} further improved on photo-z quality relative to previous COSMOS catalogues \citep{Capak07,Ilbert09,Ibert13_catalog}. The accuracy of the photo-zs has been verified by comparing them to a large number of highly reliable (97\% confidence) spectroscopic redshifts \citep{zCOSMOS_lilly07} that are available in the COSMOS field. For the objects of magnitude $i^+_{\rm AB} <22.5$ and redshift range of $z=0$--$1.2$, the photo-zs have an r.m.s scatter of $\sigma = 0.7\%$ with respect to the spectroscopic redshifts, while the occurrence of ``catastrophic failures'' with relative errors $|z_p - z_s |/(1 + z_s) > 0.15$ is only 0.51\%. For this work, we choose the median of photo-z likelihood distribution from the template fitting (``ZPDF'' in the catalogue) as the base redshift. From this redshift we calculate angular-diameter and luminosity distances, and corresponding luminosities and projected separations, assuming all galaxies follow the Hubble flow. In the process of template fitting and photo-zs estimation, \citet{Laigle16} also calculated stellar masses and star formation rates for the galaxy samples, which will be used in our analysis below. Specifically, we use ``MASS\_MED''  and ``SFR\_MED'', the medians of the stellar mass and star-formation-rate probability distribution functions (PDFs). 

\subsection{Additional spectroscopic redshifts}
We can further improve on our distance estimates by supplementing the COSMOS photo-zs with spectroscopic redshifts, where these are available. While there is no single public spectroscopic redshift catalogue for the whole COSMOS field, most of the measured redshifts in the region are now accessible through the NASA Extragalactic Database\footnote{https://ned.ipac.caltech.edu}. In addition to these redshifts, we also obtained a few other unpublished redshifts from the COSMOS collaboration (M. Salvato, private communication). The redshifts used in this work will be mainly photo-zs from the COSMOS 2015 catalogue, but replaced with spectroscopic redshifts where possible. Given the numerous literature sources and slightly different qualities of the spectroscopic redshifts, a universal redshift uncertainty of 0.0001 is assigned to each galaxy whose photometric redshift is replaced with a spectroscopic value. Absolute magnitudes and stellar masses for those objects are also corrected, based on the resulting change in the distance modulus. 

\subsection{The base sample}
\begin{figure}
\centering
 \includegraphics[width=0.99\columnwidth]{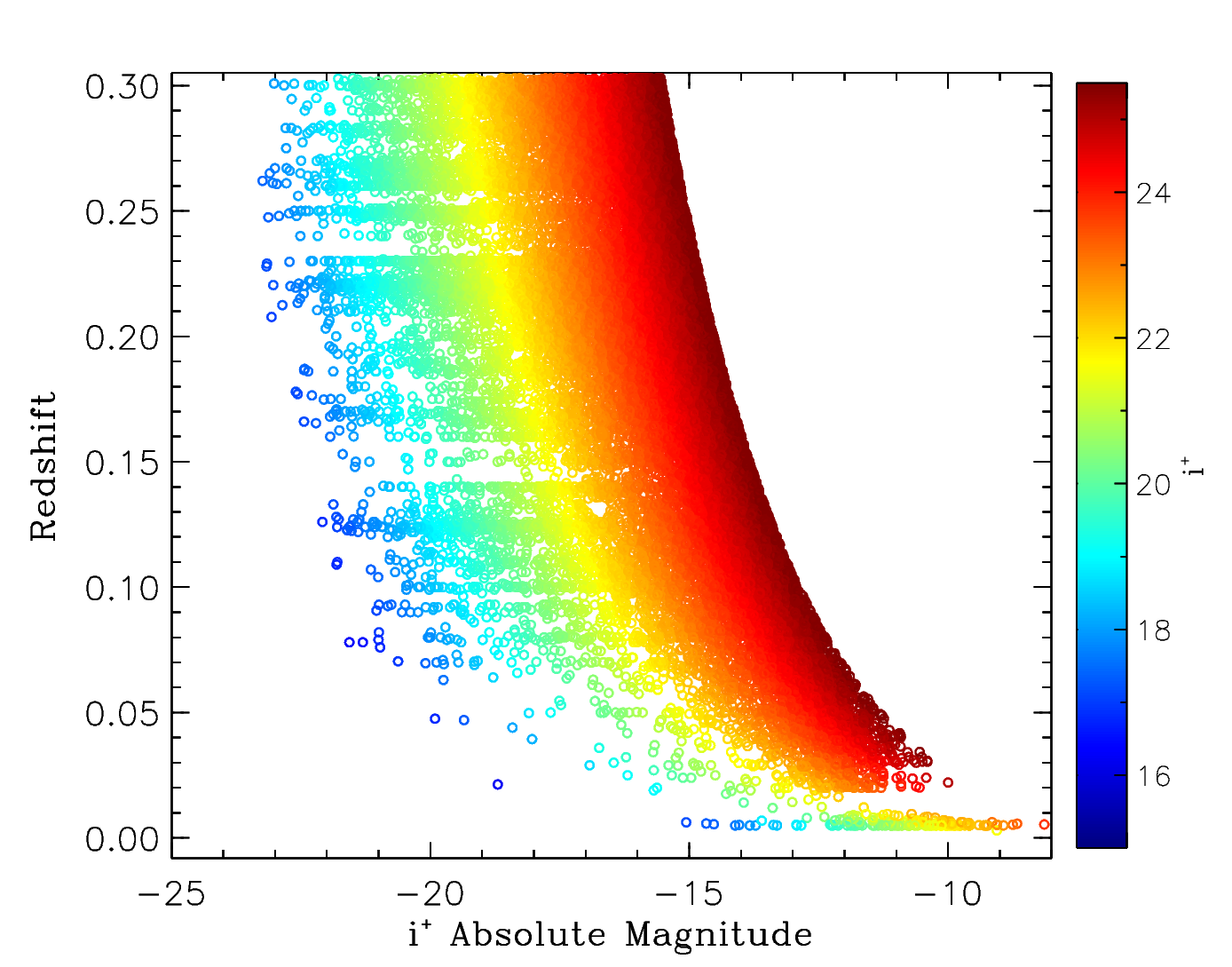}
\caption{The redshift versus i$^+$ band absolute magnitude distribution of our base catalogue, coloured by the i$^+$ band apparent magnitude, as indicated in the right-hand colour scale.}
\label{fig:z_vs_abmag}
\end{figure}

As mentioned above, the depth of the COSMOS 2015 catalogue varies across the field, depending on whether the new ``ultra-deep'' (UltraVISTA DR 2) imaging is available or not. In general, the catalogue appears to be relatively complete down to a magnitude of $i^+ < 25.5$ (MAG\_AUTO), but becomes incomplete beyond this. [The 3$\sigma$ depths in the i$^+$ band are 26.2 and 26.9 for 3\arcsec and 2\arcsec apertures respectively \citep{Laigle16}.] To ensure reasonable completeness over the redshift range of interest, we apply the following initial cuts on the catalogue, which are the same cuts used in \citet{Xi18}:
\begin{itemize}
\item $i^+ <25.5$ 
\item $0<z_{\rm pdf}<6.9$
\item $z-2\sigma_z <0.3$ 
\item $\sigma_z <0.5$ 
\end{itemize}
where $z_{\rm pdf}$ refers to the median of photo-z likelihood distribution measured using galaxy template fitting, and $\sigma_z$ refers to the photo-z error, estimated by using the 68\% confidence level upper and lower limits of the photo-z likelihood distribution provided in the catalogue (i.e.~$\sigma_z = (z_{\rm pdf}^{U68} - z_{\rm pdf}^{L68})/2$). Note that we include the broad redshift cut $0<z_{\rm pdf}<6.9$ to exclude stars and X-ray sources in the catalogue, as well as objects in the masked regions, as these objects do not have robust $z_{\rm pdf}$ estimates; we include the upper limit cut of redshift $z-2\sigma_z <0.3$ to focus on the local volume while keeping a reasonable completeness over a target redshift range of 0--0.25; finally, we use a redshift error cut $\sigma_z <0.5$ to exclude those galaxies with poor quality redshifts (mainly faint galaxies) from the further analysis. These cuts produce a base catalogue of 41,559 galaxies (37,578 after excluding galaxies with large 
redshift errors). Fig.~\ref{fig:z_vs_abmag} shows the redshift versus i$^+$ absolute magnitude distribution for our base catalogue after applying the cuts above. Given our cut in apparent magnitude, the sample galaxies have absolute magnitudes between -24 and -10 for the redshift range (z=0--0.25) we will consider below.

 \subsection{Stellar Mass Completeness}

\citet{Laigle16} estimated the stellar mass completeness of their catalogue; for redshift range $0<z<0.35$, they suggested a 90\% completeness limit of $M_*=10^{8.6} M_\odot$. We are considering systems at a slightly lower redshift range z < 0.25. Examining the stellar mass function and stellar mass errors for this redshift range, we conclude that we are complete down to at least  $M_* < 10^{8.2} M_\odot$, where the differential mass function peaks, and cut the main sample at this value. For very low redshifts ($z < 0.07$) we appear to be complete down as low as $M_* < 10^{7.2}$--$10^{7.5} M_\odot$; we will discuss local satellite abundance at these lower stellar masses below and in subsequent work. 
 
%%%% Section 3 %%%%

\section{Determining the Clustering Scale}
\label{sec:clustering scale}
We assume a model in which the galaxy with the largest stellar mass is the dominant galaxy within each halo, and resides at or close to, its geometric and dynamical centre. To tell whether a given galaxy is the dominant central galaxy (CG, or ``primary'' hereafter) within its own halo, or a potential satellite of another primary, we need to search its surroundings to see if there is another more massive galaxy close by. If there are more massive neighbours nearby, this suggests the galaxy may be satellite, whereas if all nearby galaxies are less massive, it suggests the galaxy is a primary. To quantify the characteristic scale on which satellites are associated with their primaries, we measured the clustering pattern of all the galaxy pairs in our base sample. Based on these clustering results, we will quantify ``spatially nearby'' and specify the region of interest (ROI) for the primary-secondary classification.  

\subsection{Halo Mass Assignment} 

We expect the characteristic extent of the satellite distribution in a halo to scale with its virial radius. We calculate a fiducial halo mass and virial radius for each galaxy, assuming that it is the CG of its host halo. These masses and virial radii will be used to characterize the clustering throughout this work. 

To estimate halo masses, we could in principle use abundance matching, assuming a monotonic relation between stellar mass and halo mass that we derived empirically by comparing the observed stellar mass function and the predicted halo mass function within the observation volume. However, the effective area of the COSMOS field (after correcting for masking as discussed below) is only 1.46 deg$^2$, giving an effective comoving volume of 1.67$\times 10^5$ Mpc$^3$ up to $z=0.25$. The cosmic variance in the mean density for a volume this size is a factor of approximately 0.4 \citep{Somerville04}; if we consider ten independent, equal-volume redshift slices, the relative cosmic variance of each slice increases to 0.7. Thus, there is a large systematic uncertainty in the normalization of the halo mass function within this volume. Instead, we use the Stellar-to-Halo Mass Relation (SHMR) derived by \citet{Behroozi13} (B13 herafter). They provide a formula (Eqn.~\ref{eqn:behrooz13_3a}) for the inverse Halo-to-Stellar Mass Ratio (HSMR), with parameters as listed in Appendix \ref{apd:shmr_bias}.

We note, however, that a combination of observational errors in the stellar mass estimates, intrinsic scatter in the SHMR, and the non-linearity of the halo mass function will bias halo masses estimated directly from the B13 formula. As there are many more low mass halos than high mass ones, errors in stellar mass will more frequently scatter objects into a given mass range from below than from above. Thus, directly applying the HSMR to estimate the halo mass corresponding to an observed stellar mass will lead to systematic overestimates, especially at the high-mass end. To quantify and correct this bias, we have performed a Monte-Carlo simulation of the effect of errors to obtain a bias-corrected HSMR based on [B13]. We explain this correction in more detail in Appendix \ref{apd:shmr_bias}.

Given an estimate of the halo mass for each galaxy, we also assign a corresponding virial radius and virial velocity. The virial radius is taken to be $\left(\frac{3 M_h}{4 \pi \rho_c \Delta_c} \right)^{1/3}$, where $\rho_c$ is the critical density of the universe and $\Delta_c$ is the mean overdensity of the halo within the virial radius, with respect to the critical density. We use the fitting formula from \citet{Bryan1998} for $\Delta_c$. 
Once the virial radius has been calculated, the virial velocity is given by $\sqrt{G M_h/R_{\rm vir}}$.

The stellar mass completeness limit is estimated empirically, following the method described in \citet{Laigle16}. They calculated the  $K_s$-band magnitude limits for the COSMOS catalogue to be 24.0 and 24.7 for the deep and ultra-deep fields respectively. In this work, we choose a limit of 24.0 in order to have uniform depth across the whole field. Given this magnitude limit, the limiting mass a galaxy would need to have to be observed at a given redshift is calculated as:
\begin{equation}
\log M_{\rm lim} = \log M - 0.4 \left( K_{s{\rm lim}} -K_s \right)
\end{equation}
Next, a stellar mass limit is estimated for each redshift bin, within which 90\% of the galaxies lie, given the stellar mass errors. We also calculate a corresponding halo mass limit in each redshift bin using the bias-corrected HSMR (although in this case we ignore the effect of scatter on the completeness threshold). 

\subsection{Defining a Region of Interest (ROI)}
\label{sec:ROI_define}
Around each CG, we define a ``Region of Interest" (ROI) in which we will search for potential satellites. The size of ROI is determined by two considerations: first, the clustering signal should be consistent across systems with different CG masses, and second, the ROI should include most of the ``one-halo'' clustering signal associated with the main halo around the CG, while excluding the regions that are dominated by the ``two-halo'' or background terms. We have explored different possible choices of ROI boundaries by measuring the two-dimensional (line-of-sight and projected) clustering signal of all pairs in our base catalogue. 

For each pair in the catalogue, the galaxy with the larger stellar mass is assumed to be the primary. We then count pairs as a function of separation in projected distance and line-of-sight velocity offset, scaled by the halo virial radius and the velocity error respectively. These scaling choices are found to give us relatively consistent clustering signals for primaries of different masses (as shown in Fig.~\ref{fig:rv_phase_density} below). Note that some fraction of an annulus around a given primary may be missing from the catalogue, as it overlaps with field boundaries and masked regions. We carefully measure the shape of the survey boundaries and masked regions to produce a single template for the whole field. Monte Carlo sampling of this template is then used to determine the area completeness $\eta$ around each primary as a function of radius, as explained in Appendix \ref{apd:masking}, and the counts are corrected by this factor.

Fig.~\ref{fig:rv_phase_density} shows the density of pair counts in 2D phase space for all galaxies (top panel), and binned by primary halo mass (bottom panels), where the primary is defined to be the member of the pair with the larger stellar mass. A clear overdensity of pairs can be seen both in the projected separation and in the velocity separation directions. Overall, primaries with larger masses show a stronger clustering signal. At the same time, the clustering patterns in the different mass ranges have a similar dependence on separation scale: they all have the strongest clustering within 0.5 $R_{\rm vir}$. The signals all extend to fairly large radii, but start to drop significantly after 1.5--2 $R_{\rm vir}$. Along the velocity axis, which is scaled by redshift error, the signals in all three mass bins drop off at a similar rate, reaching the background level at $\Delta v\sim 1$--1.5 $\sigma_{\Delta v}$. It is worth noting the slight asymmetry of the clustering signal along the velocity separation axis, with slightly more negative velocity separations than positive ones. This is due to the incompleteness at the faint, high-redshift end of the survey volume (as shown in Fig.~\ref{fig:z_vs_abmag}). We calculate the velocity offset with respect to the more massive (and thus more luminous) member of the pair, which as a result of Eddington bias due to incompleteness, trends to be further away on average. Thus, it produces a negative velocity offset more often than a positive one. 

\begin{figure}
\centering
\includegraphics[width=\columnwidth]
{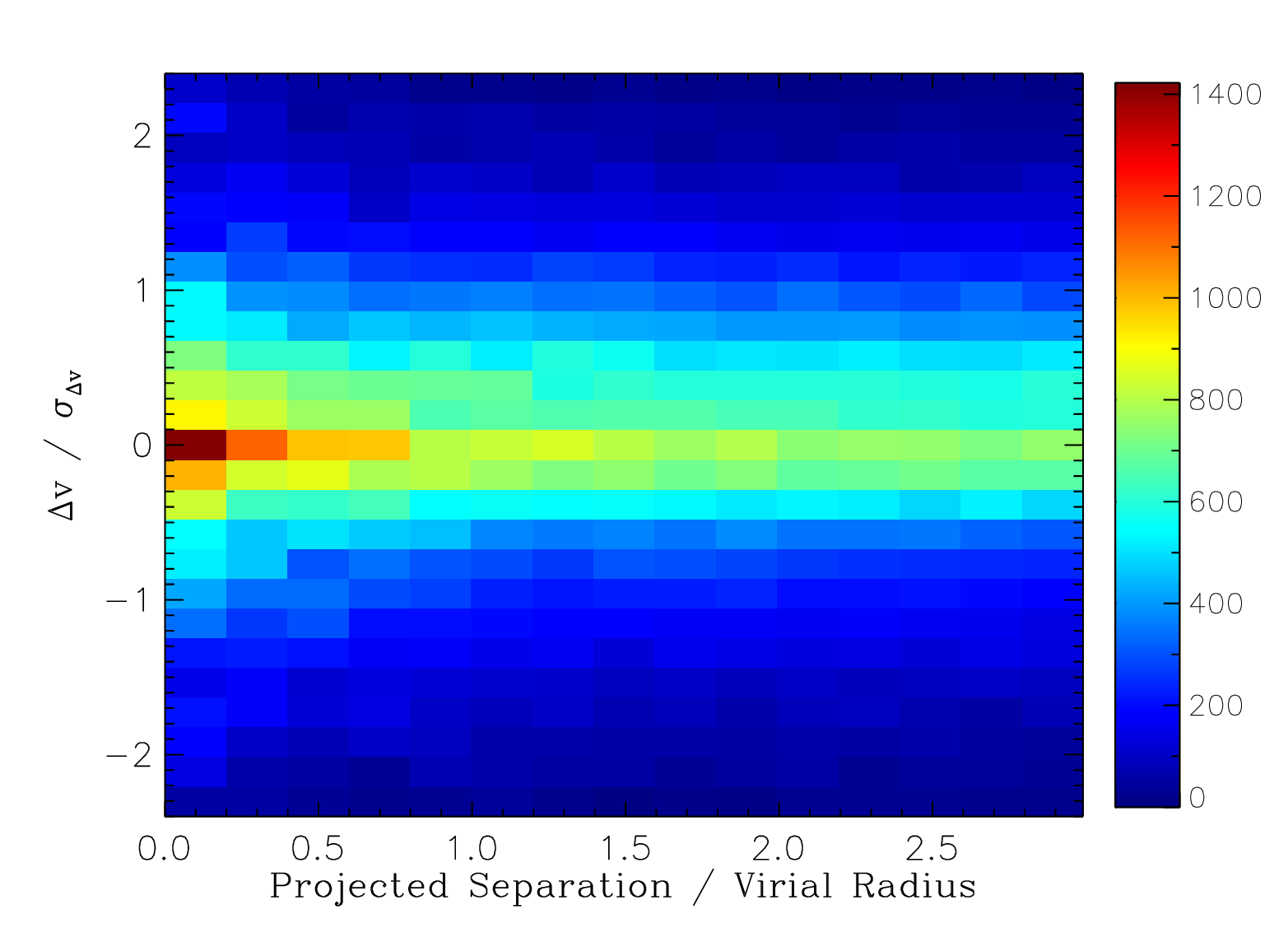}
\includegraphics[width=\columnwidth]
{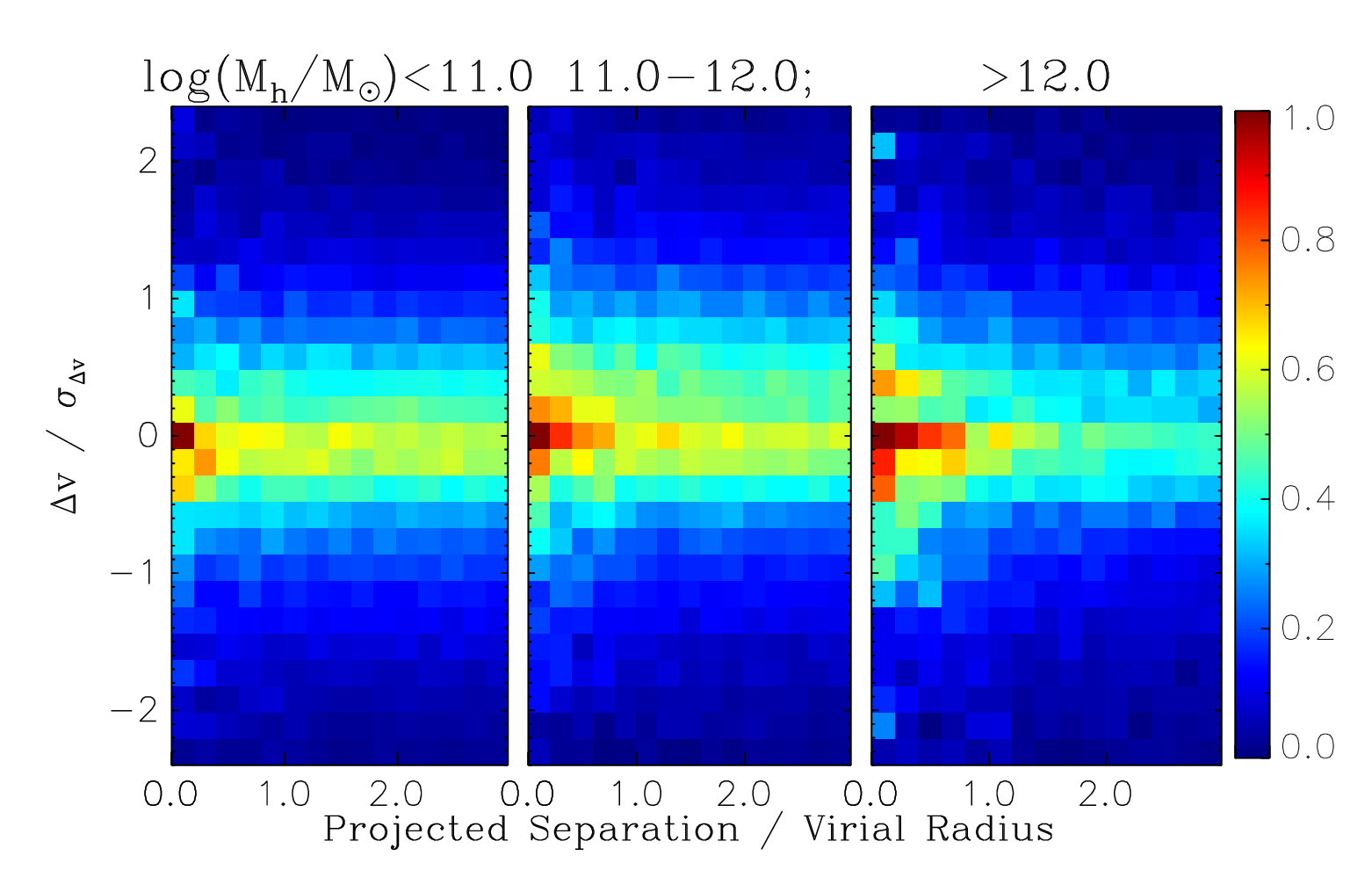}
\caption{Primary-secondary clustering signal in the base catalogue. The colour scale shows the density of galaxy pairs $\Sigma = N/\Delta R\Delta v$ as a function of projected separation and velocity offset, where these have been scaled by the estimated virial radius and velocity error, respectively. Top panel: all primaries; bottom panels: same quantity normalized and binned by primary mass.}
\label{fig:rv_phase_density}
\end{figure}

Based on these clustering patterns, around each primary galaxy we will define a ROI for potential satellites using the following cuts:
\begin{enumerate}
\item A cut in projected separation (assuming both galaxies are at the line-of-sight distance of the primary), scaled by the virial radius of the primary: $R_P/R_{\rm vir} < A $
\item A cut in velocity difference, scaled by the circular velocity of the primary: $\Delta v/v_{\rm vir} < B $ 
\item A cut in velocity difference relative to the uncertainty in velocity difference between the primary and secondary: $\Delta v/\sigma_{\Delta v} < C $ 
\end{enumerate}
Note the final cut depends on the secondary properties as well as those of the primary, so this is determined for each galaxy pair individually. 
Any secondary galaxy in a pair that meets conditions (i) and ((ii) or (iii)) is considered a potential satellite of the more massive member. We will choose the values ($A, B, C$) = (3.0, 2.0, 1.5) as our default, but test the effect of changing these definitions of the ROI in Section.~\ref{sec:test} below.

%%%% Section 4 %%%%

\section{The iterative clustering method}
\label{sec:p-s selection}

Our goal is to quantify satellite abundance using the strength of the clustering signal. A obvious complication is that any given galaxy may be a central or a satellite; without further selection, the `raw' clustering signal measured in the previous section consists of a complicated sum of central-satellite and offset-satellite terms \citep[e.g.][]{Leauthaud2012}. Depending on the stellar mass and range of separations considered, the latter can significantly bias estimates of the true satellite abundance. 

To avoid this complication, our strategy will be to identify the galaxies most likely to be centrals, and use the clustering signal around these objects as a template to separate out central and satellite contributions. In this section, we will first describe the initial, `first-run' primary selection; then we will model the clustering of secondaries around these first-run primaries to determine primary and secondary probabilities for all galaxies in the sample, and finally we will test the results of iterating over this process, by adding new potential primaries to our initial sample, weighted by their primary probability, and remeasuring a weighted clustering signal around the enlarged primary sample.

\subsection{Initial primary selection}
\label{sec:primary_algorithm}

We want to select primaries that dominate a ROI that scales with their halo mass, as described in the previous section. Since the ROI is larger for the more massive systems, a smaller system that has no more massive companions within its own ROI can still lie within the ROI of a larger system. This asymmetry naturally leads us to a hierarchical search algorithm, where we start searching around the most massive systems first. The detailed steps are as follows:
\begin{enumerate}
\item All galaxies in the catalogue are assigned a halo mass and virial velocity dispersion based on their stellar mass, as if they were the CG in their own host halo.
\item We then go through the catalogue in ranked order of stellar mass, selecting the most massive galaxy in the catalogue as the first primary.
\item All galaxies that lie in the ROI around the first primary (as defined in the previous section) are classified as its secondaries and removed from the list of potential primaries;
\item The next most massive unclassified galaxy is then selected as the next primary candidate.
\item We check the stellar masses of all galaxies within the ROI around this next candidate. If the candidate is the most massive galaxy in its ROI, then it is classified as a primary and the other galaxies in the ROI are classified as its secondaries. 
\item We iterate over the last two steps until all the galaxies in the catalogue are classified as primaries or secondaries.
\end{enumerate}

This produces our first run primary sample. To guarantee reasonable completeness, we make two additional cuts on this initial sample:
 \begin{enumerate}
\item Primaries with ROIs that are heavily affected by survey boundaries or masking are removed from the initial primary list.
\item Primaries with redshifts higher than 0.25 are removed from the initial primary list. 
\end{enumerate}
These cuts are necessary to remove primaries close to the sample boundaries, either on the sky or in redshift. These galaxies may have more massive companions that lie just outside the field or beyond our redshift cut. Thus, there is a higher probability that they are not actual CGs, but are in fact satellites of another, more massive galaxy. 

Overall, this process is very conservative in selecting primaries, producing a sample of 1,490 galaxies that is incomplete (in the sense of missing many genuine CGs), but relatively uncontaminated by satellites. Cutting out systems with redshifts exceeding 0.25 reduces the number of first-run primaries to 873, while excluding those with area completeness less than 0.65 (i.e. those with ROIs that are masked or cut off by field boundaries by more than 35\%), reduces the number to 815. 

\subsection{Clustering of the First-run Primary and Secondary Samples}

To study satellite abundance and its dependence on primary properties, we first need to model and separate the contributions to the clustering signal from the satellite population and the background galaxy population. Having classified potential primaries and secondaries, we measure the surface number density of secondaries within the ROI and in an extended region around it (with the same velocity offset limits, but extended out to 3.2 R$_{\rm vir}$ in order to have a better estimate of the background surface density).

Around each primary, we count secondaries in annuli spaced evenly in $\log[R_P/R_{\rm vir}]$. The annuli range from 0.1--3.2 $R_{\rm vir}$, in steps of 0.25 dex. We exclude secondaries with projected separations of less than 0.1 $R_{\rm vir}$ ($\sim$25 kpc, for the Milky Way) to avoid outlying HII regions or other components of the central galaxy. We calculate the surface density of secondary galaxies around each primary in physical units (Mpc$^{-2}$), assuming all secondaries lie at the same distance as the primary. Given counts $N^{i,j}$ in radial bin $i$ of projected area $A_i$ around primary $j$, the surface density is:
\begin{equation}
\Sigma_{i,j} =\frac{ N^{i,j}  } {\eta_i A_i} =\frac{ N^{i,j}  } {\eta_i 2\pi R^i \Delta R} 
\label{eqn: sd_rrvir}
\end{equation}
where $\eta_i$ is the mean area completeness in radial bin $i$ (estimated as described in Appendix \ref{apd:masking}), $R^i$ is the mean radius of the bin, and $\Delta R$ is the width of the bin. As we want to scale the surface density of secondary galaxies in units of $R_{\rm vir}$, it is useful to define a separation variable $X_i \equiv R_P^{i,j} /R_{\rm vir}^j$. Thus, Eqn.~\ref{eqn: sd_rrvir} can be written as:
\begin{equation}
\Sigma_{i,j} =\frac{ N^{i,j} } {\eta_i 2\pi (X_i \cdot R_{\rm vir}^j) \Delta X \cdot R_{\rm vir}^j}
= \frac{ N^{i,j} } 
{\eta_i 2\pi  X_i \Delta X \left( R_{\rm vir}^j\right)^2} ~.
\end{equation}
In what follows, we will fit the surface density in bins of primary redshift and mass. Where the primary sample contains more than 5 objects, we use the bootstrap method to estimate the uncertainties in the surface densities, by re-sampling the primary sample 120 times. As the bootstrap method does not work well when the sample size is smaller than 5, we have also calculated Poisson uncertainties for each bin. The final uncertainties for the bin are taken to be the larger of the two values.

\begin{figure}
\centering
\includegraphics[width=\linewidth]{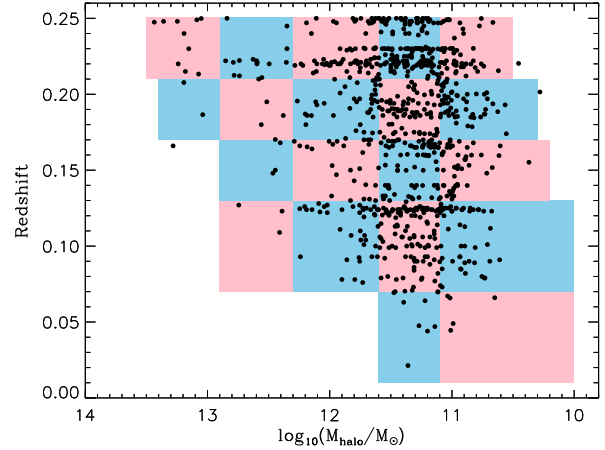}
\caption{The redshift-halo mass distribution of the first-run primary sample. The coloured boxes show the boundaries for the 20 subsets used to fit the mass and redshift dependence of the clustering signal.}
\label{fig:primary_split}
\end{figure}

The secondary surface density consists of two main parts: the contribution from clustered satellites and the contribution from background or foreground galaxies.
\begin{equation}
\Sigma(X| M_{\rm halo}, z) = \Sigma_{\rm sat} (X| M_{\rm halo}) + \Sigma_{\rm bg} 
\end{equation}
The first component $\Sigma_{\rm sat}$ should correlate with the halo mass of the primary, but should be roughly independent of redshift over the narrow redshift range considered here, while the second component should be roughly independent of halo mass, but should depend on redshift. There should also be a more extended clustered component due to large-scale structure (the "two-halo" term, e.g. \citet{Cooray2002} for a detailed review), but the characteristic scale of this component (cf. 4--8 Mpc) is much larger than the scales considered here. Thus, we treat it as a constant with respect to radius, and include it in the foreground/background term. To fit the two terms, we split the primaries into 5 fixed redshift slices and 2--5 halo mass bins per slice, with adaptive boundaries as shown in Fig.~\ref{fig:primary_split}. 

We assume that the satellite distribution roughly matches the subhalo distribution, which in turn approximately follows a Navarro-Frenk-White (NFW --  \citet{NFW96}) density profile. Over the range of radii we are most sensitive to, a projected NFW profile scales as $r^{-2}$ in the outer parts of the halo, and is somewhat shallower in the inner parts. We could fit the density profile of the satellite component with the exact form of a projected NFW profile, assuming some mean concentration-mass relation. Given the low SNR of the satellite component, however, we choose to fit it with the simplified form:
\begin{equation}
\label{eq:fit_profile}
\Sigma_{\rm sat}(X) 
=S_{\rm halo} F(X)
= \frac{S_{\rm halo}}
{X^2 + \alpha X +\beta} 
\end{equation}
where $S_{\rm halo}$ is an overall normalization (in units of Mpc$^{-2}$) that is independent of $X$ but depends on the primary halo mass, while 
$\alpha$ and $\beta$ are parameters describing the radial dependence. We have compared this simplified form to a projected NFW profile over the range X = 0.05--3, for concentration parameters c= 5--15, and found that using fixed values $\alpha = 0.2$ and $\beta = 0$ gives a good fit in all cases. For this choice of parameters, $\int_0^1 2\pi F(X) X dX = \log[1+\alpha] - \log[\alpha] = 11.18$, so the total number of satellites within the virial radius is $N_{\rm sat} = 11.18\,S_{\rm halo} R_{\rm vir}^2$. 

\begin{figure*}
\centering
\includegraphics[width=1.95\columnwidth]
{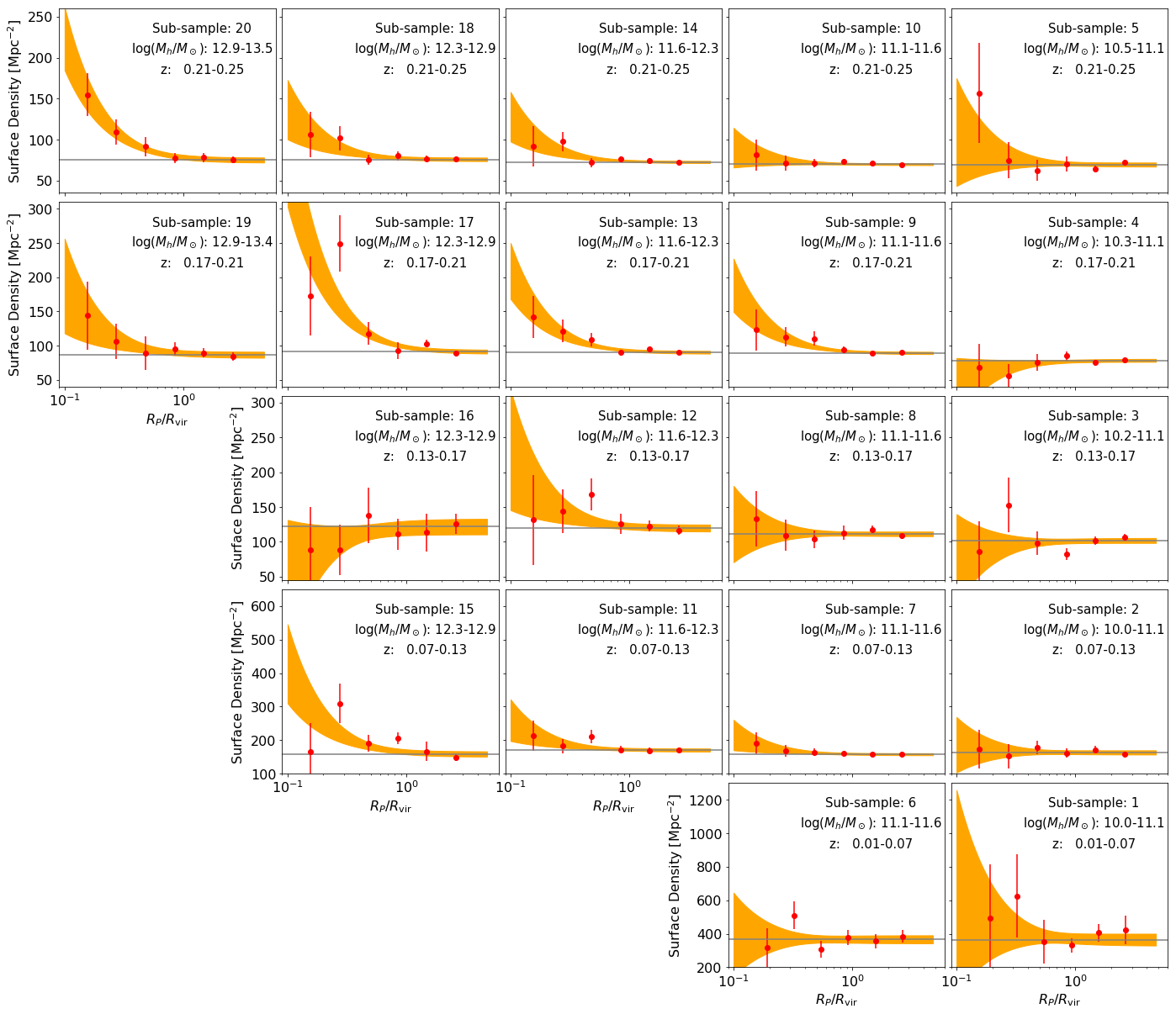}
\caption{
The surface density of secondaries around each of the primary subsamples shown in Fig.~\ref{fig:primary_split} (separate panels, with redshift increasing from bottom to top and halo mass increasing from right to left). Red points with error bars show the measured surface number density in radial bins, while the orange shading shows the 1-$\sigma$ region around the best fit from Eq.~\ref{eq:fit_profile}. The horizontal grey lines indicate the background level in the best fit model.}
\label{fig:sd_20panel}
\end{figure*}

The resulting fits for the secondary surface density (fitting $S_{\rm halo}$ and $\Sigma_{\rm bg}$ jointly for each individual primary bin, with $\alpha, \beta$ set to fixed values (0.2, 0)) are shown in Fig.~\ref{fig:sd_20panel}. Summing over all primary bins, a clustered excess in $\Sigma$ is detected at a SNR of approximately 11. We can see that the overall background surface density (as measured in physical units) decreases with primary redshift, while the clustered satellite component increases with primary mass. In the next section, we will explore these correlations in more detail. 

\subsubsection{Satellite/halo component}

\begin{figure}
\centering
\includegraphics[width=\columnwidth]{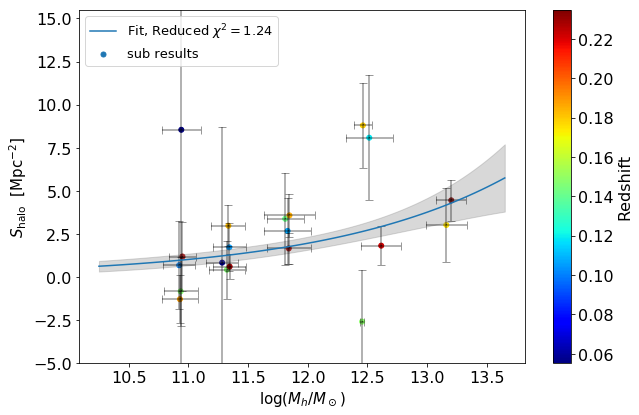}
\caption{Average $S_{\rm halo}$ versus halo mass ($\log_{10}( M_{\rm h}/M_\odot)$), with colour indicating the mean redshift of the primary sub-sample. The blue curve and grey shaded area show the best-fit model of the mass dependence (Eqn.~\ref{eq:fitmassdpdc}), together with the 1-$\sigma$ uncertainty range.}
\label{fig:fit_mp_sh_mh}
\end{figure}

\begin{figure}
\centering
\includegraphics[width=\columnwidth]{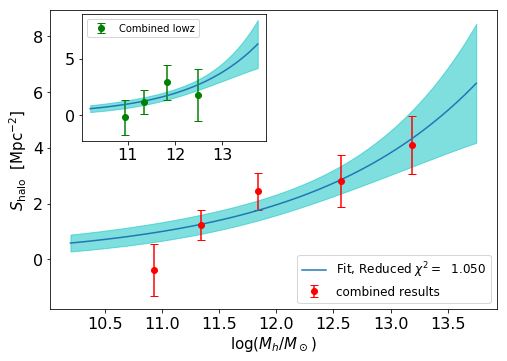}
\caption{As Fig.\ref{fig:fit_mp_sh_mh}, but with sub-samples of similar mean halo mass combined into single bins for clarity. The top-left subplot shows the results for primaries with redshift $z<0.15$. }
\label{fig:fit_sh_mh_combo}
\end{figure}

We fit the halo mass dependence of $S_{\rm halo}$ over the 20 primary bins using a linear relation in log-log space:
\begin{equation}
\label{eq:fitmassdpdc}
 \log_{10}\left( \frac{S_{\rm halo}}{1 {\rm Mpc}^{-2}} \right) = a\cdot \log_{10} \left(\frac{M_{\rm halo} }{ 10^{12} M_\odot} \right)+b
\end{equation}
where $a$ and $b$ are free parameters. We choose $10^{12} M_\odot$ as the pivot mass in our fit, as this is roughly the median halo mass of our 20 primary bins. The projected area of a halo will scale as the virial radius squared, that is as $M^{2/3}$. If we assume systems have a fixed number of satellites per unit halo mass (as expected from subhalo abundance, e.g.~ \citealt{Gao+2004}), then the projected surface density should go as $M^{1/3}$, so we expect $a\sim0.33$. The value of $b$ (the normalization at $M_{\rm halo} = 10^{12} M_\odot$) will depend on the depth of the catalogue, as discussed below. From our fits, we find $a=0.30^{+0.11}_{-0.10}$ and $b=0.26^{+0.08}_{-0.11}$, so the scaling with halo mass seems fairly consistent with the expected value.

Fig.~\ref{fig:fit_mp_sh_mh} shows our fit for $S_{\rm halo}$ as a function of mean halo mass, over all 20 bins in primary mass and redshift. Given its sensitivity to smaller radial bins with larger errors, the fitted value of the parameter has a SNR $\sim 8$, significantly lower than the SNR for the whole clustering signal. 
To illustrate the dependence on halo mass more clearly, we also show in Fig.~\ref{fig:fit_sh_mh_combo} a version combining bins with similar mean halo masses. 
We note that at the low-mass end ($\log(M_h/M_\odot) < 11$), the fitted value is actually negative (although consistent with zero, given the uncertainties). This may be partly due to completeness problems at low mass, which affect the high redshift bins in particular. Repeating the fitting process for low redshift primaries (z<0.15) only, we obtain a less negative value, that is once again consistent with zero.
(as shown in the sub-panel of Fig.~\ref{fig:fit_sh_mh_combo}). 

%---------------------------------------------
\subsubsection{Background component}

\begin{figure}
\centering
\includegraphics[width=\columnwidth]{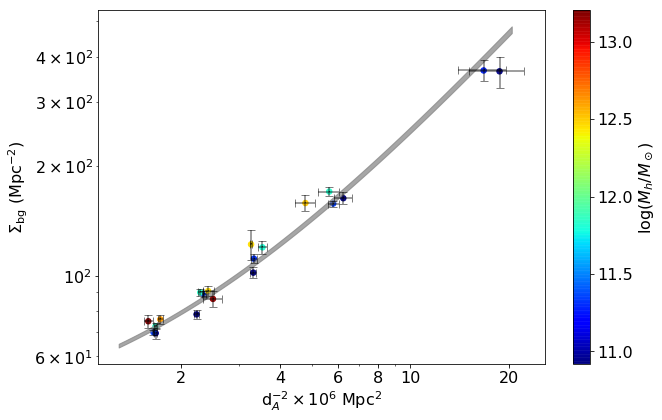}
\caption{The background surface density $\Sigma_{\rm bg}$ versus ${d_A}^{-2}$. The grey area shows the 1-$\sigma$ uncertainty range around the best-fit model (Eqn.~\ref{eq:fitbg}).}
\label{fig:fit_mp_sbg_da2}
\end{figure}

Our surface densities are calculated in physical units (Mpc$^{-2}$) at the distance of the primary. Thus, if the foreground/background component consisted of a fixed field population with a broad redshift distribution and thus a fixed number per square degree, we would expect its inferred physical surface density to scale as ${d_A}^{-2}$, where $d_A$ is the angular diameter distance of the primary. (It is worth noting that this assumption will not necessarily hold if extending the method to higher redshift.) Fig.~\ref{fig:fit_mp_sbg_da2} shows the fitted value of the background surface density $\Sigma_{\rm bg}$ in each bin, versus the average value of ${d_A}^{-2}$ for that bin. We fit the trend with a simple linear model:
\begin{equation}
\label{eq:fitbg}
\Sigma_{\rm bg} = c + d \cdot 
\left[ \left(
\frac{d_A}{10^3 {\rm Mpc}} 
\right)^{-2}-2.0
\right]
\end{equation}
where $c$ and $d$ are free parameters. 

Fitting the 20 bins gives tight constraints on the parameters: $c = 79.5^{+0.6}_{-0.6}$ and $d = 21.6^{+0.5}_{-0.5}$. Note that if the background scaled exactly as the inverse of the angular diameter distance, we would expect the constant term $c-2d \sim 0$ to be small; in practice, various minor effects, notably the varying width of the redshift range $\Delta z$ over which we measure the background, will cause the background density to deviate from the simple scaling. As it is, for our fitted values of the parameters we find $c-2d = 36.3$, which is small relative to the typical values of $\Sigma_{\rm bg}$.

\subsubsection{Single-step Fit}

\begin{figure}
\centering
\includegraphics[width=\columnwidth]{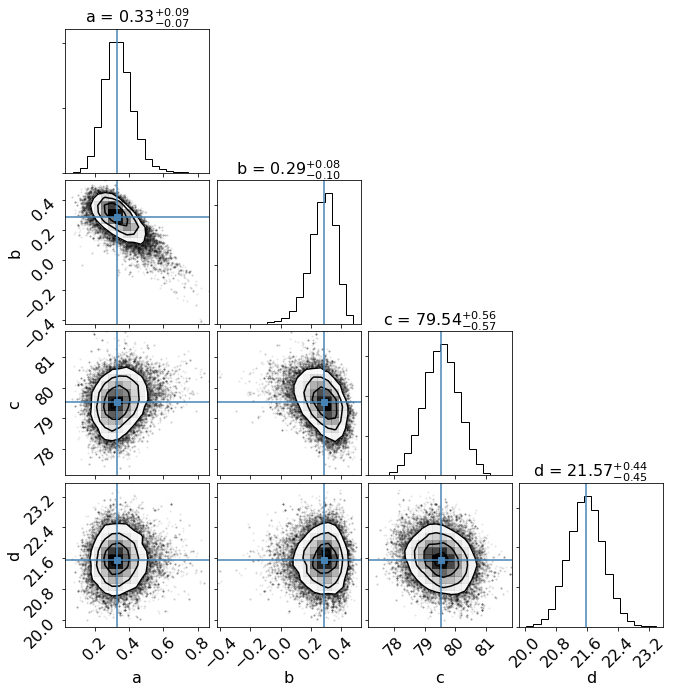}
\caption{Likelihood distributions for the clustering model parameter values, derived by fitting the full model to all 20 primary sub-samples simultaneously. Panels show the distribution marginalized over the two (or, on the diagonal, all three) other model parameters.}
\label{fig:params4_corner}
\end{figure}

While the two-step fitting procedure outlined above is useful to illustrate the features the model, it is more robust to fit the entire 4-parameter model for both terms in the surface density in a single step, given the potential covariance between the model parameters. We use the function Minimizer.emcee\footnote{
See \href{https://lmfit.github.io/lmfit-py/fitting.html}{this page} for a detailed description of LMFIT and Minimizer.emcee } from the Python module LMFIT (Least-Squares Minimization Fitting) to do MCMC sampling of the likelihood \citep{Foreman-Mackey2013-emcee}. The marginalized results of this fit are shown in Fig.~\ref{fig:params4_corner} and are also included in Table~\ref{Tab:fit_results}. We note that the parameters $a$ and $b$ are strongly (anti-)correlated; indeed, with higher SNR data we could imagine a more detailed, HOD-based fit to the halo-mass dependence of the satellite abundance. There is also some correlation between parameters $c$ and $b$ (or to a lesser degree $c$ and $a$), indicating that satellite abundance estimates do require careful accounting for the background term.

\subsubsection{Assigning Probability}
Given our model fit to the clustering measurements, we can estimate the amplitude and radial distribution of the satellite component and background components around each primary. We define the probability of a secondary galaxy in the ROI being an actual satellite as:
\begin{equation}
P^{\rm sat}_{i,j}(X, M_h, z^p) 
= \frac{\Sigma_{\rm sat}}{\Sigma_{\rm tot}} 
= \frac{\Sigma_{\rm sat} (X, M_h)}{\Sigma_{\rm sat}(X, M_h) + \Sigma_{\rm bg} (z^p)}\,,
\label{eq:prob_define}
\end{equation}
where $M_h$ and $z^p$ are the halo mass and redshift of the primary, respectively.

This equation can result in very small probabilities at large radii. Since real satellites (objects that have crossed the virial radius at least once) almost all lie within 3 $R_{\rm vir}$ \citep[e.g.][]{Wetzel2014}, we truncate the probability around $X=3$ as follows:
\begin{equation}
P'(X) = P(X) \cdot \frac{1}{1+ 1000^{X-3}} ~.
\label{eq:prob_cutoff}
\end{equation}

 \subsection{Iterating over the Fit}

Our initial primary selection ignores many galaxies that may well be primaries, but appear close to more massive systems when seen in projection. To attempt to correct for this, we iterate over our clustering measurement, including a weighted contribution from all galaxies, proportional to their probability of being a primary.

For each iteration, we run the top-down selection again. During the new selection process, every galaxy starts with a 100\%\ probability of being a primary. Starting with the most massive galaxy as the first primary, we assign probabilities of nearby galaxies being its satellites, using the method described in the last section. The probability of each of these galaxies being independent primaries is reduced accordingly. We then proceed through the catalogue in order of decreasing stellar mass. If a galaxy has a probability of being a primary between zero and 1, we estimate that nearby galaxies have a probability of being its satellites that is the {\it product} of its probability of being a primary and the satellite probability given in section 4.2.4. Proceeding through the entire catalogue in decreasing order of stellar mass ensures that the calculation is well-defined and that every galaxy is assigned a final probability of being a primary, equal to 100\%\ minus the sum of all probabilities that it is a satellite of nearby systems.

Running through this process once, we find 8,920 primaries with probabilities greater than 0.99, of which 3,567 lie at redshifts of 0.25 or less. If we further remove systems with areal completeness of 0.65 or less, the number of high-confidence primaries is 3,246, versus 815 in our original sample; repeating this exercise for systems with primary probabilities of 0.999 or more, reduces the number to 1,478.

If we include primaries with probabilities greater than 0.99 (and weight all satellites by their CGs primary probability), the effect of iteration on the fit to the clustering signal is shown in Table~\ref{Tab:fit_results}. Overall, the parameter values after iteration show good consistency with our initial estimates, shifting by less than 2-$\sigma$ in all cases. The uncertainties in the fitted parameter values drop, but only slightly. This suggests that in a dataset like the COSMOS catalogue that has extremely precise photometric redshifts, even the first-run sample of isolated primaries can provide a good estimate of satellite abundance. Since the use of lower probability primaries may dilute the clustering signal and introduce some bias, in what follows we will use our initial, first-run estimates of the fitted parameters to derive satellite abundance. We anticipate that in datasets with less accurate redshifts, iteration will become more important in deriving accurate estimates of the clustering signal. 

\begin{table}
\caption{The model fitting results from the first run and first iteration}
\begin{tabular}{c|c|c|c|c}
\hline \hline
&  a & b& c & d \\ \hline 
F0  
&$0.33^{+0.09}_{-0.07}$  &$0.29^{+0.08}_{-0.10}$ 
&$79.5^{+0.6}_{-0.6}$&$21.6^{+0.4}_{-0.5}$ \\ \hline
F1  &$0.26^{+0.07}_{-0.07}$  &$0.38^{+0.05}_{-0.06}$ 
&$79.3^{+0.5}_{-0.5}$&$20.8^{+0.4}_{-0.4}$  \\

\hline
\end{tabular}

\label{Tab:fit_results}
\end{table}

%%%% Section 5 %%%%

\section{Estimating Satellite Abundance}
\label{sec:results}

In this section, we will make some basic estimates of the overall abundance of satellites, as well as their abundance as a function of properties such as stellar mass or luminosity. In each case, our estimate is based on the clustering signal, which is typically small compared to the background. The simplest way to estimate satellite abundance is to count every galaxy in the ROI, weighted by the satellite probability calculated in Section 4.2.4, so we will use this approach first in Section 5.1, referring to it as ``method A''. 

As explained below, method A assumes that the clustering amplitude is uncorrelated with stellar mass, luminosity, or any of the other secondary properties considered. More generally, we expect the fraction of galaxies in the ROI that are true satellites to depend on these other properties. In section 5.2 we develop a more sophisticated approach, ``method B'', similar to the one introduced in \cite{ST14}, that attempts to correct for possible correlations in the limit of a weak clustering signal. Future surveys with stronger detections of clustering should be able to bypass these complications by dividing the galaxy sample directly into bins of secondary property value before they measure the clustering amplitude, simplifying the analysis considerably; we call this ``method C''.

\subsection{Abundance Estimates Using Method A}
\label{subsec:methodA}

In method A, to estimate satellite abundance $N^i_{\rm sat}$ around primary galaxy $i$ we simply add up the probabilities $P_j^i$ of each galaxy $j$ in its ROI being a true satellite:
\begin{equation}
N_{\rm sat}^{i} = \sum_j P_j^i ~.
\end{equation}
Note that although $P_j^i$ can remain non-zero at large distances from the primary, to compare to previous results from literature we set a radial limit of 1.5 $R_{\rm vir}$ by default, and only count towards the total satellite abundance secondaries that lie within this projected separation of the primary.

\begin{figure}
\centering
\includegraphics[width=0.9\columnwidth]{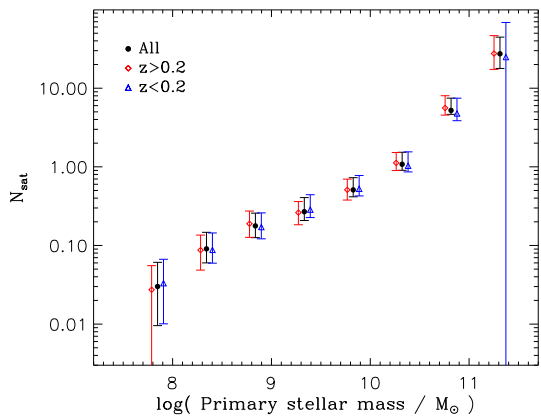}
\caption{Satellite abundance as a function of primary stellar mass, estimated using method A.
Red and blue points show the results for independent primary samples at higher redshifts ($z=$0.17--0.25) and lower redshifts ($z=$0.07--0.17), respectively.}
\label{fig:nsat_ms_sct}
\end{figure}

We can study the dependence of satellite abundance on primary mass by stacking systems with similar stellar or halo masses, as shown in Fig.~\ref{fig:nsat_ms_sct} (black points with error bars). As expected, there is a strong trend in satellite abundance with halo mass. As a consistency check, we also split our primary sample in two by redshift, and calculate satellite abundance separately for each of the two sub-samples (red and blue points). The results for both sub-samples show good consistency with those for the whole sample. We note that $N_{\rm sat}$ increases faster with primary stellar mass at the high mass end of the range ($M_{\rm h} > 10^{10.5} M_\odot$). This is consistent with the pattern seen in Halo Occupation Distribution (HOD) modelling \citep[e.g.][]{Seljak2000,Peacock2000,Berlind2002} and is a result of the changing slope of the SHMR; halo mass increases faster with stellar mass at large stellar masses, resulting in a faster increase in satellite numbers.

\subsection{Abundance Estimates Using Method B}

Method A provides the simplest estimate of satellite abundance, and is guaranteed to be correct when averaging over all galaxies in the secondary sample used to calculate the satellite probability (Eqns.~\ref{eq:prob_define} \& \ref{eq:prob_cutoff}). It is not necessarily correct, however, for subsamples of secondaries selected by luminosity, stellar mass, colour, or other properties, if these properties are correlated with the clustering amplitude. As a simple example, one can imagine a galaxy population that consisted of two distinct types, labelled ``red" and ``blue". If the red galaxies were completely clustered, but the blue galaxies were completely unclustered, then we would measure some intermediate average clustering strength for the combined population, and give every galaxy a satellite probability based on this average value. If we weighted all galaxies by this average satellite probability, but then split them back into subsamples by colour, we would conclude that the satellite and background populations both had the same net colour distribution. In effect, the true colour distribution of the satellite population (100\%\ red, in this example) would be contaminated by the colour distribution of the field population. More generally, whenever the background population is significantly different from the satellite population, the satellite properties inferred using method A will be biased towards those of the background population.

With sufficient SNR in the clustering signal, we could avoid this problem by selecting subsamples with a limited range of the desired satellite property (stellar mass, luminosity, colour, etc.) {\it before} measuring the clustering amplitude and calculating the satellite probability (we refer to this as ``method C''). In the limit of low SNR, however, splitting the galaxies into narrow bins in a given property will increase the shot noise in the background estimate until it is unacceptably large. Instead, we have developed an intermediate solution, ``method B'', based on the approach in \citet{ST14}.

In method B, first the number of the background galaxies within a given radius is estimated for each primary, by summing up the non-satellite probabilities of each pair:
\begin{equation}
N_{\rm bg}^i = \sum_j (1 -P_j^i) ~.
\end{equation}
We then measure the fraction of {\it all} background galaxies (i.e.~summing over the ROIs of all primaries, or over the whole field) with a property of interest (e.g. stellar mass, luminosity, colour, etc.) in a given range, and scale the total number of background galaxies in the ROI by this fraction. This gives the expected background contribution to a particular subsample, that we then subtract to calculate satellite abundance. For instance, if we want to measure the luminosity function of satellites,  $\Phi_{\rm sat}$, we first need to measure the total luminosity function within the ROI, $\Phi_{\rm TTL}$, and the total luminosity function for all background galaxies, $\Phi_{\rm bg}$. Then we remove the background contribution from the total abundance in each bin $k$ in luminosity, such that:
\begin{equation}
\Phi_{{\rm sat},k}= \Phi_{{\rm TTL},k} - \frac{N_{\rm bg}}{N_{\rm TTL}} \Phi_{{\rm bg},k} ~.
\end{equation}
Here $N_{\rm bg}$ and $N_{\rm TTL}$ refer to the number of background galaxies, and the total number of galaxies within the radial cut around each primary, respectively (before any cut in luminosity). Note that this approach can be used for individual primaries, except for the lowest-mass systems, where the galaxy counts are so small that Poisson fluctuations dominate. To correct for this, if the number of objects within the ROI is less than three times the number of luminosity bins, then we stack results for multiple primaries at similar redshifts, and use the average signal.

\begin{figure}
\centering
\includegraphics[width=0.91\columnwidth]{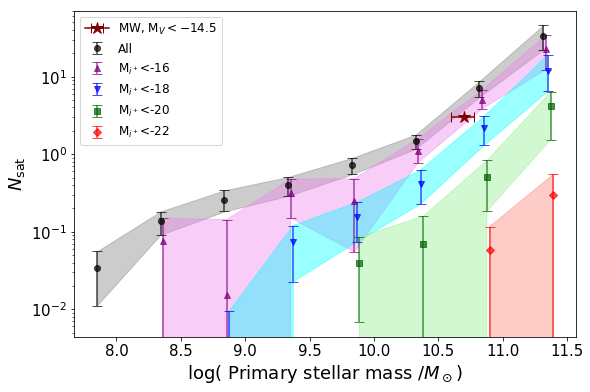}
\includegraphics[width=0.91\columnwidth]{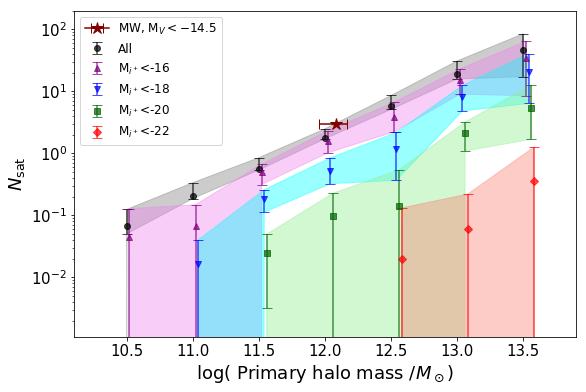}
\caption{
Satellite abundance as a function of primary stellar/halo mass (top/bottom panels respectively), estimated using method B. The individual shaded regions show results for different cuts in satellite luminosity, as indicated in the legend. The brown diamond shows the number of MW satellites brighter than $V = -14.5$ with a galactocentric distance greater than 20 kpc (three satellites meet these criteria).}
\label{fig:nsat_msmh_abmag2}
\end{figure}

We used this method to calculate satellite abundance for different luminosity ranges (below), as well as full satellite luminosity functions (see Sec.~\ref{sec:LF_abmf}). Fig.~\ref{fig:nsat_msmh_abmag2} shows abundance for various cuts in 
$M_{i^+}$. Overall, the dependence of abundance on primary stellar (top panel) or halo (bottom panel) mass has a similar form for different luminosity cuts, although there may be a truncation at lower stellar masses that depends on the luminosity cut. Here too, this pattern is very similar to those seen for brighter galaxies in HOD modelling. Plotted as a function of stellar mass, satellite abundance shows a change in slope around 10.5 $M_\odot$ at all luminosities. Plotting as a function of halo mass, this feature disappears, confirming that it is a result of the non-linear SHMR. We also indicate on the plot the abundance of MW satellites brighter than $V = -14.5$ with a galactocentric distance greater than 20 kpc (brown diamond -- three satellites meet these criteria), assuming a MW halo mass of $12.1 M_\odot$, as discussed below.

\subsection{Comparison to Previous Work}

\begin{figure}
\centering
\includegraphics[width=0.9\linewidth]{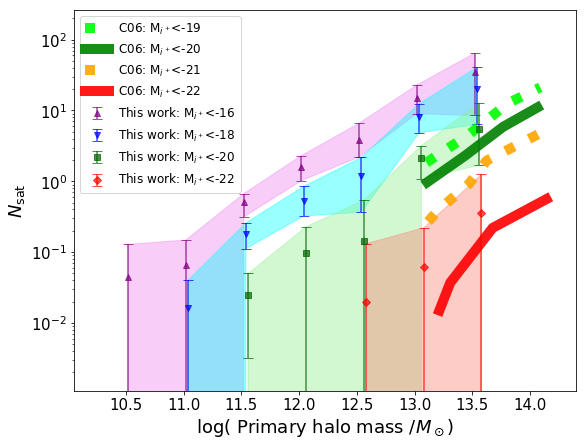}
\caption{
Comparison to the HOD-based results of \citet[Fig.~5]{Conroy2006}. Their luminosity bins have been converted to $i^+$-band magnitudes, assuming a mean colour of $\langle M_r-M_{i^+}\rangle = 0.25$, and have had 1 subtracted from them to account for the central galaxy. 
}
\label{fig:compare_conroy2006}
\end{figure}

For massive galaxies, a number of other estimates of satellite abundance exist in the literature. In Fig.~\ref{fig:compare_conroy2006}, we compare our abundance estimates to the results of \citet[C06]{Conroy2006}. These are based on HOD modelling \citep[e.g.][]{Seljak2000, Peacock2000, Zehavi2002} of the luminosity functions and correlation functions of samples from the SDSS \citep{SDSS} and DEEP2 \citep{DEEP2} surveys. HOD modelling provides an estimate of the average number $\langle N_{\rm gal}\rangle$ of galaxies within a halo of a given mass, so in the limit where $\langle N_{\rm gal}\rangle \gg 1$,  $\langle N_{\rm gal}\rangle - 1$ should match our measured satellite abundance. As $\langle N_{\rm gal} \rangle$ decreases, some halos will contain no galaxies over a given magnitude limit, so we can only compare results in the large $\langle N_{\rm gal}\rangle$ regime, i.e. for large halo masses. In addition, the results in C06 are given in bins of $M_r -5\log(h)$; we convert to our $i^+$-band assuming a fixed mean colour $\langle M_r-M_{i^+} \rangle \sim 0.25$ (roughly the value measured for our sample in the COSMOS catalogue), which produces a shift of $(M_r-M_{i^+}) - 5\log(h)\sim 1$ magnitude exactly in the C06 bin boundaries. 

Given these conversions, examining Fig.~\ref{fig:compare_conroy2006}, we see that there is excellent agreement between our results and those of C06. Our estimated satellite abundance matches that measured by C06 to within half a standard deviation, for all four of the magnitude cuts in that study (two lie between our luminosity cuts, but are clearly consistent with our results.) The slope of the $N_{\rm sat}$--$M_{\rm h}$ relation is harder to judge given the limited baseline in C06, but generally it appears to be consistent with our inferred slope at halo masses of $\log(M/M_\odot)> 13$ or more. The agreement between these two sets of results is particularly striking, given that they employ completely different samples and methods, and that there is no parametric freedom in adjusting our results. Overall, SDSS provides a more robust estimate of satellite abundance for massive halos, but as a deeper survey with more accurate redshift estimates, COSMOS is better able to probe the low-halo-mass regime.

\begin{figure}
\centering
\includegraphics[width=0.9\columnwidth]{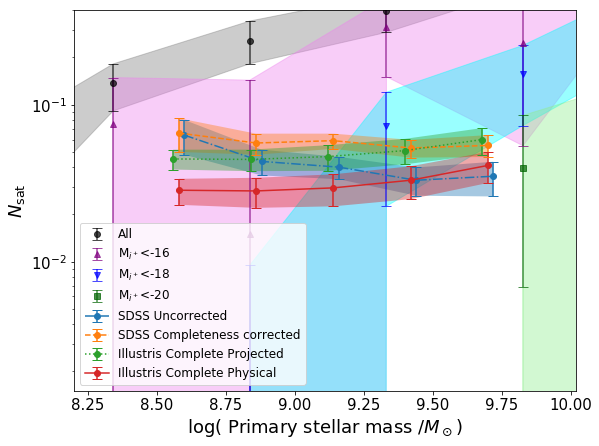}
\caption{Comparison to the observed and mock results of  \citet[][; see their Fig.~11,12]{Besla2018}. 
(Note that the points are shifted to left or right by up to 0.02 dex for clarity).}
\label{fig:compare_besla2018}
\end{figure}

We have also compared our results to more recent work by \citet[B18]{Besla2018}, which is one of the few studies to estimate satellite abundance at lower halo mass. Fig.~\ref{fig:compare_besla2018} compares our results to theirs, for the primary stellar mass range $10^8$ -- $10^9 M_\odot$. The B18 results are based on a SDSS spectroscopic sample at redshift 0.013--0.0252, with $r$-band magnitudes between 14 and 17.77, and primary stellar masses in the range 0.2--5 $\times 10^9 M_\odot$. (The apparent magnitude limit of 17.77 corresponds to absolute magnitude limits of -16.03 or -17.5 at z=0.013 or z=0.0252 respectively, equivalent to -16.3 and -17.8 in the $i^+$-band.) 
They also compare these to a mock catalogue generated using the Illustris hydrodynamical simulations \citep{Vogelsberger2014, Nelson2015}. Four sets of results are shown. The "uncorrected SDSS" results are from raw counts of nearby companions; the ``completeness corrected'' version is after correcting for observational selection effects, using the mock catalogues. The ``physical" simulation results show the abundance of real satellites, while the "projected" counts show the result including a background contribution introduced by projection effects. Completeness and projection corrections move the measured abundance up and down, respectively; overall, the best estimate of satellite abundance from B18 is the completeness corrected (orange) curve, decreased by a factor of $\sim$50\% to account for projection effects.

Over the mass range 0.2--5 $\times 10^9 M_\odot$ covered by B18, our estimates for magnitude cuts $M_{i^+} < -16$ or $M_{i^+} < -18$ are consistent with their measured values. Given their effective magnitude limits range from -16.3 and -17.8 in the $i^+$-band, there once again seems to be excellent consistency in the overall abundance estimated by the two methods. In contrast, B18 measure almost no trend in satellite abundance with primary stellar mass. It is worth pointing out, however, that the mean redshift of the sample in B18 is lower at smaller primary masses (see their Fig.~3). Thus, the satellite luminosity function may be measured to greater depth for these systems, explaining the flatter slope. B18 also point out that they may be biased towards preferentially identifying multiple systems at low stellar mass, due to their bluer colours. We conclude that our results are consistent with B18, once again despite very different methods and samples.

%-----------------------------------------------------------------------
\subsection{The Satellite Luminosity Function}
\label{sec:LF_abmf}

Finally, we can use method B to estimate satellite luminosity functions directly (for comparison, results using method A are shown in Appendix \ref{apd:method AvsB}). In Fig.~\ref{fig:abmf_3mh_B} we present the luminosity function of satellites for three subsets of primaries. The subsets were chosen such that the mean halo mass of the middle bin, $\langle M_{\rm h} \rangle= 12.1 M_\odot$, is close to the estimated mass of the Milky Way (MW) or M31 \citep{Bland-Hawthorn2016, Posti2019}, such that we can compare to the observed luminosity functions for these systems. For method B, we are able to measure the satellite luminosity functions reliably down to absolute magnitudes of -14. Within this magnitude range, the observed abundance of satellites around the MW and M31 are close to the average value. One exception is at the bright end of the MW satellite luminosity function, where the presence of the LMC and SMC represent a slight (1--2 $\sigma$) departure from the average. This unusual feature of the MW's satellite population has been noted and discussed extensively elsewhere (e.g. \citealt{Robotham2012} -- see \citealt{ST14} for further references). We will consider satellite luminosity and stellar mass functions in more detail in a subsequent paper.

\begin{figure}
\centering
\includegraphics[width=\columnwidth]{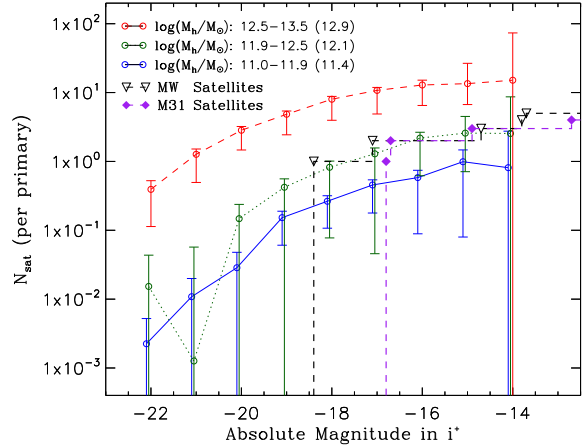}
\caption{
Satellite luminosity functions, for three ranges of primary halo mass, estimated using method B (see Appendix \ref{apd:method AvsB} for the same results derived using method A). The mean halo mass for each range is given in parentheses. The observed luminosity functions of the  MW and M31 are shown for comparison (black and purple points and lines). These magnitudes have been converted from the $V$-band, assuming a fixed average colour of $V - i^{+} = 0.3$. 
}
\label{fig:abmf_3mh_B}
\end{figure}

%%%% Section 6 %%%%

\section{Testing for Systematic Uncertainties in the Method}
\label{sec:test}

While we have shown that our method produces estimates of satellite abundance consistent with previous studies using larger samples, there remain a number of choices, assumptions or free parameters in the method that could take on different values. In this section, we will perform a set of tests to understand the effects of the various assumptions and free parameters in the clustering method. 

\subsection{Null Test}
\begin{figure}
\centering
\includegraphics[width=0.9\linewidth]{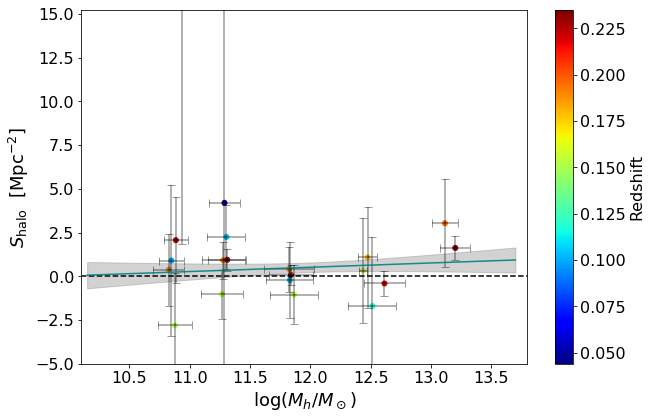}
\caption{Results of a null test where the halo component of the clustering signal is measured using galaxies outside the velocity limits of the ROI. A black dashed line corresponding to no clustering is included for comparison.}
\label{fig:null_test}
\end{figure}

First, as a null test, we calculated the clustering signals around our initial sample of primaries, using only secondaries that lay within the radial cuts, but outside the redshift cuts we defined in Section \ref{sec:ROI_define}. Following the procedure in Section \ref{sec:p-s selection}, we measured the surface number density of the secondaries for the 20 bins in primary mass and redshift, and re-fit our surface density model. The satellite component from the fit is shown in Fig.~\ref{fig:null_test}. 

Most of secondaries in this test should not be physically associated with the primaries, expect for a small number of real satellites that are scattered into the foreground or background by the redshift errors. Thus, we expect the clustering signal to be close to zero, and to show little dependence on the primary halo mass. This is confirmed in Fig.~\ref{fig:null_test}.

\subsection{Parameter and Systematic Tests}
We have tested for systematic effects and uncertainties in our method by varying the parameters that define the initial primary selection and the calculation of the satellite probability. The tests include:
\begin{itemize}
\item Using radial cuts of 2 $R_{\rm vir}$ or 3.5 $R_{\rm vir}$ to define the radial extent of the ROI.

\item Using velocity cuts of 1 $\sigma_{\Delta v}$ or 3 $\sigma_{\Delta v}$ to define the velocity extent of the ROI.

\item Increasing/decreasing the stellar mass by 0.16 dex, which is comparable to or larger than the typical stellar mass uncertainties in our data.

\item Increasing/decreasing the halo mass derived from our fiducial SHMR by 50\%. 

\item Varying the slope of bias-corrected SHMR (see Appendix \ref{apd:shmr_bias}) at the high mass end to 1.5 and to 2.5, with respect the original value $\sim$2.1 from B13.

\item Varying the definition of the virial radius, increasing or decreasing it by 20\%. 

\item Adding 1$\sigma$ scatter to the initial stellar masses before ranking them. (We perform this test three times to check the consistency of the potential effects.)

\item Keeping the primary-secondary selection fixed, but adding 0.2 $R_{\rm vir}$ scatter to the coordinates of the primaries, to test the effects on potential mis-centering. (We repeat this test three times to check for the consistency of the effects.)

\end{itemize}
In each case, all other parameters and steps in the method are kept fixed. The results of these tests are summarized in Tab.~\ref{Tab:test_results}.

\begin{table}
\caption{The fitting results of all tests }
\begingroup
\renewcommand{\arraystretch}{1.25} 
\begin{tabular}{c|c|c|c|c}
\hline \hline
&  a & b& c & d \\ \hline 
F0  
&$0.33^{+0.09}_{-0.08}$  &$0.29^{+0.08}_{-0.10}$ 
&$79.5^{+0.6}_{-0.6}$&$21.6^{+0.5}_{-0.5}$ \\ \hline
LMh 
&$0.29^{+0.08}_{-0.07}$  &$0.24^{+0.09}_{-0.10}$ 
&$79.0^{+0.5}_{-0.5}$ &$21.7^{+0.5}_{-0.5}$ \\
SMh 
&$0.28^{+0.08}_{-0.07}$  &$0.37^{+0.06}_{-0.08}$ 
&$79.0^{+0.5}_{-0.6}$&$21.6^{+0.5}_{-0.5}$ \\
LMs 
&$0.30^{+0.09}_{-0.08}$  &$0.12^{+0.09}_{-0.12}$ 
&$78.8^{+0.5}_{-0.5}$ &$21.7^{+0.5}_{-0.5}$ \\
SMs 
&$0.45^{+0.12}_{-0.10}$  &$0.33^{+0.08}_{-0.09}$ 
&$80.2^{+0.6}_{-0.6}$&$22.0^{+0.5}_{-0.5}$ \\ \hline
$\sigma_{\Delta v}$ 1 
&$0.32^{+0.08}_{-0.07}$  &$0.22^{+0.07}_{-0.09}$ 
&${\bf 78.2}^{+0.5}_{-0.5}$&$21.2^{+0.4}_{-0.4}$ \\
$\sigma_{\Delta v}$ 3
&$0.21^{+0.10}_{-0.09}$
&$0.29^{+0.08}_{-0.10}$
&$79.9^{+0.7}_{-0.7}$
&$21.1^{+0.6}_{-0.5}$  \\ \hline
R2  
&$0.30^{+0.06}_{-0.06}$ 
&$0.32^{+0.06}_{-0.06}$ 
&$78.7^{+0.5}_{-0.5}$&$21.4^{+0.4}_{-0.4}$ \\
R3.5  
&$0.23^{+0.09}_{-0.08}$ 
&$0.28^{+0.06}_{-0.08}$ 
& ${\bf 82.2}^{+0.5}_{-0.5}$&$21.0^{+0.4}_{-0.4}$ \\ \hline
LR  
&$0.28^{+0.09}_{-0.08}$  
&${0.14}^{+0.08}_{-0.10}$ 
&${\bf 78.1}^{+0.5}_{-0.5}$ 
&$21.2^{+0.4}_{-0.4}$ \\
SR  
&$0.26^{+0.10}_{-0.09}$ 
&${0.28}^{+0.09}_{-0.12}$ 
&${80.7}^{+0.7}_{-0.7}$ 
&$22.1^{+0.5}_{-0.5}$ \\ \hline
H SHMR 
&$0.20^{+0.06}_{-0.06}$  
&$0.31^{+0.05}_{-0.07}$ 
&$80.2^{+0.5}_{-0.5}$ 
&$21.3^{+0.3}_{-0.4}$ \\
L SHMR 
&$0.34^{+0.10}_{-0.09}$  
&$0.28^{+0.07}_{-0.09}$ 
&$79.4^{+0.6}_{-0.6}$ &$21.6^{+0.5}_{-0.5}$ \\ \hline
RS 1 
&$0.26^{+0.08}_{-0.07}$
&$0.28^{+0.07}_{-0.09}$
&$78.6^{+0.6}_{-0.5}$
&$21.6^{+0.5}_{-0.5}$\\
RS 2 
&$0.25^{+0.08}_{-0.07}$
&$0.29^{+0.07}_{-0.08}$
&$78.7^{+0.6}_{-0.5}$
&$21.6^{+0.5}_{-0.5}$\\
RS 3
&$0.25^{+0.08}_{-0.07}$
&$0.28^{+0.07}_{-0.09}$
&$78.6^{+0.5}_{-0.5}$
&$21.6^{+0.5}_{-0.6}$\\ \hline
CS 1 
&$0.34^{+0.12}_{-0.10}$
&${\bf 0.05}^{+0.11}_{-0.15}$
&$79.4^{+0.5}_{-0.5}$
&$21.8^{+0.5}_{-0.5}$\\
CS 2 
&$0.25^{+0.10}_{-0.08}$
&$0.11^{+0.9}_{-0.12}$
&$79.3^{+0.5}_{-0.5}$
&$21.8^{+0.5}_{-0.5}$\\
CS 3
&$0.26^{+0.08}_{-0.08}$
&$0.21^{+0.08}_{-0.10}$
&$79.6^{+0.6}_{-0.6}$
&$21.6^{+0.5}_{-0.5}$\\
 \hline

\end{tabular}

\label{Tab:test_results}
\endgroup
(Key to the tests: \\
 F0 = original fit result; \\
 LMh = larger halo mass; \\ 
 SMh = smaller halo mass; \\
 LMs = larger stellar mass; \\
 SMs = smaller stellar mass; \\
 $\sigma_{\Delta v}$ 1 = using 1$\sigma_{\Delta v}$ velocity cut to define ROI; \\
 $\sigma_{\Delta v}$ 2 = using 3$\sigma_{\Delta v}$ velocity cut for ROI; \\
 R2 = using 2 $R_{\rm vir}$ projected separation cut for ROI; \\
 R3.5 = using 3.5 $R_{\rm vir}$ projected separation cut for ROI; \\
 LR = larger $R_{\rm vir}$ (increasing the primary virial radius by 20\%);\\
 SR = smaller $R_{\rm vir}$ (decreasing the primary virial radius by 20\%);\\
 H SHMR = using a higher slope of HSMR at high mass end;  \\
 L SHMR = using a lower slope of HSMR at high mass end; \\
 RS = ranking shuffle (adding 1-$\sigma$ scatter to the primary stellar masses before ranking them); \\
 CS = centering shift (adding 0.2 $R_{\rm vir}$ scatter to the coordinates of the primaries before measuring clustering).
 ) 
 \end{table}

\subsection{Discussion}
Considering the results of the individual tests in detail, changing the halo mass mainly just shifts the points horizontally on the $S_{\rm halo}$--$M_h$ plot, so the fitted parameters $a$ and $b$ remain relatively constant. Changing the stellar mass has quite a different effect, however. As the HSMR is shallow at the low-mass end but steep at the high-mass end, increasing or decreasing the stellar mass does not change the inferred halo mass much at the low-mass end, but can produce significant change at the high-mass end. As a result, the slope ($a$) of the $S_{\rm halo}$--$M_h$ relation is more strongly affected. 

Similarly, changing the slope of the HSMR at the high-mass end will mainly affect the halo mass estimates in this range. As a result, the slope $a$ is shifted systematically to higher or lower values. As for the mass ranking test, in addition to the random scatter in individual mass estimates and resulting variations in detailed primary selection, there is a net change in the mass function. Since there are more low-mass galaxies than high-mass ones, adding the random scatter tends to increase the number of massive galaxies relative to the fiducial catalogue. This leads to slight shifts in the fitted satellite abundance, although they are less important than in the case of the mass tests. 

The parameter values obtained in each test are given in Tab.~\ref{Tab:test_results}. Tests that produce a variation of more than 2-$\sigma$ in the fitted parameters are highlighted in bold. We note that only one test (our first re-centering test) produces a significant change in the parameters of the satellite component. Three of the 38 tests produce significant deviations in the background fit, but this is only slightly higher than the expected rate of 2-$\sigma$ deviations given the random errors (8\% versus 5\%). Thus overall, the systematic uncertainties associated with our tests do not appear to significantly increase the random errors quoted in the fiducial model fit. 

%%%% Section 7 %%%%

\section{Conclusions}
\label{sec:conclusion}

In this work, we have developed and tested a method for quantifying satellite abundance, using galaxy clustering. The method establishes a basic template for the radial dependence and amplitude of the satellite component of the clustering signal by using a subsample of isolated (or at least locally dominant) systems, but then applies this template iteratively to estimate the probability that any given galaxy in the field is a satellite of a nearby system.  (Note that the form of the template assumes that the surface number density of background galaxies is inversely proportional to the square of angular-diameter distance; this assumption works well at low redshift, but may need modification if applying the method at higher redshift.) In that sense it is similar to crowded-field photometry, where an initial sample of isolated stars is used to determine the point spread function of the image, and that point spread function is then applied iteratively to the entire field. Using our method, we have estimated satellite abundance as a function of primary stellar and halo mass, and also measured the satellite luminosity function, over a very broad range of primary halo mass (10$^{10}$--10$^{13.5} M_\odot$). We have also tested the method for systematic uncertainties by varying the model parameters, and found variations in the final results that are generally smaller than our random error estimates. 

We have compared the results of this new technique to several previous estimates of satellite abundance from the literature, that were derived using larger catalogues. Our results are fully consistent with those of \citet{Conroy2006} at the high mass end, and of \citet{Besla2018} at the low mass end, while covering a much larger range in primary mass overall. We have also compared our measured luminosity functions to those of the dominant LG galaxies, assuming an average halo mass of 12.1 $M_\odot$ for these systems. The LG satellite populations seem fairly typical, with the exception of the bright satellites of the MW (the LMC and the SMC), as noted previously in the literature. The main purpose of this paper was to describe and validate our method; in subsequent work we will consider in more detail the properties of the detected satellite populations, including their spatial distribution, colours, star formation rates, and dependence on primary properties. 

The COSMOS catalogue was chosen for this work for its deep photometry and extremely accurate photo-zs. Other deep surveys with accurate distance information will also be good candidates to apply our method. One potentially important survey is planned with SPHEREx\footnote{see this \href{official website}{https://spherex.caltech.edu/} for more details}, an all-sky survey satellite with a wide-field spectral imager. SPHEREx is currently scheduled to launch in 2024, and will produce, during its two-year mission, four all-sky maps, with hundreds of millions of near-infrared stellar and galactic spectra (0.75--5.0 micron) \citep{Bock2018_SPHEREx_short, Spangelo2015_SPHEREx_strategy}. The redshifts in the SPHEREx surveys are estimated by fitting template SEDs to observations, similarly to COSMOS. While COSMOS used photometry in 30 bands to derive its photo-zs, SPHEREx will produce low-resolution (R$\sim$20--100) spectra, with a similar final redshift accuracy, as discussed in \citet{Stickley2016_SHHEREx_redshift}. While the main survey will be shallower than COSMOS, two regions at the polar caps will be visited multiple times, providing $\sim$100 square degrees of coverage to a depth similar to COSMOS. Thus, SPHEREx should provide a redshift catalogue of similar accuracy to the COSMOS catalogue used here, but covering an area 50 times larger. The resulting increase in the SNR of the clustering signal would allow much finer binning in primary or secondary properties, giving a much more detailed view of the relationship between satellites and their central galaxies.

\section*{Acknowledgements}

We thank N.~Afshordi, A.~Broderick, M. Balogh, M.~Hudson, and our friends and collaborators from the COSMOS survey for their comments and advice. JET acknowledges support from the Natural Science and Engineering Research Council of Canada, through a Discovery Grant. The COSMOS 2015 catalog is based on data products from observations made with ESO Telescopes at the La Silla Paranal Observatory under ESO programme ID 179.A-2005 and on data products produced by TERAPIX and the Cambridge Astronomy Survey Unit on behalf of the UltraVISTA consortium. 

\section{Data Availability}
Most of the data underlying this article are publicly available. The COSMOS 2015 catalogue \citep{Laigle16} can be accessed from the COSMOS website, at \url{http://cosmos.astro.caltech.edu/page/photom}. A few spectroscopic redshifts that are unpublished from the COSMOS collaboration (M. Salvato, private communication) will be shared on reasonable request to the corresponding author with permission of the COSMOS collaboration. 
The derived data generated in this research will also be shared on reasonable request to the corresponding author.


\bibliographystyle{mnras}
\bibliography{ref}

%%%%%%%%%%%%%%%%% APPENDICES %%%%%%%%%%%%%%%%%%%%%

\appendix

\section{Bias in Halo Masses Derived from the SHMR}
\label{apd:shmr_bias}
 
Throughout this work we assume the SHMR proposed by \citet{Behroozi13}: 
\begin{equation}
\log_{10} (M_*(M_h)) = \log_{10} (\epsilon M_1) 
+ f \left(\log_{10} \left(\frac{M_h}{M_1}\right) \right)
- f(0) \,,
\label{eqn:behrooz13_3a}
\end{equation}
where the function $f(x)$ is defined as:
\begin{equation}
f(x) = -\log_{10} (10^{\alpha x} +1) + \delta 
\frac{(\log_{10}(1+\exp(x)))^\gamma}{1+\exp(10^{-x})} ~.
\end{equation}
The free parameters vary with redshift as follows:
\begin{align}
\log_{10}(M_1)&= M_{1,0} + (M_{1,a} (a-1) +M_{1,z} z) \exp(-4a^2) \\ \nonumber
\log_{10} (\epsilon) &= \epsilon_0 + (\epsilon_{a} (a-1) +
\epsilon_{z} z) \exp(-4a^2) +\epsilon_{a,2}(a-1) \\ \nonumber
\alpha&= \alpha_0 +(\alpha_a (a-1)) \exp(-4a^2) \\ \nonumber
\delta&= \delta_0 +(\delta_a (a-1) + \delta_z z) \exp(-4a^2) \\ \nonumber
\gamma&= \gamma_0 +(\gamma_a(a-1) +\gamma_z z) \exp(-4a^2) ~. \nonumber
\end{align}
where $a = 1/(1+z)$ is the scale factor. (The 1-$\sigma$ uncertainty range in these parameter values is listed on p.9 of \citet{Behroozi13}.) In our case, since our primary sample covers the fairly narrow redshift range of $z = 0$--0.25, little variation is predicted in the SHMR. Thus, we simply use an intermediate redshift of z$=0.15$ for the analysis below. 

\begin{figure}
\centering
\includegraphics[width=0.99\columnwidth]{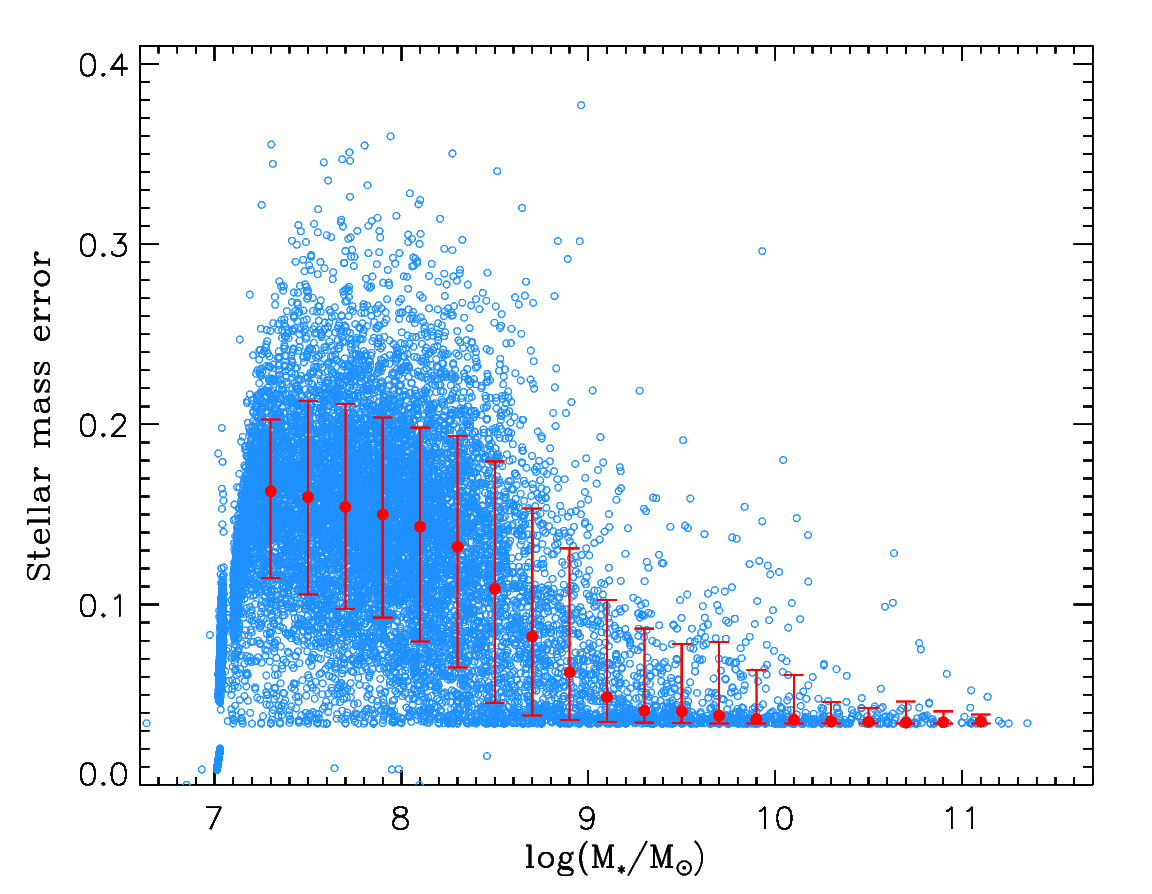}
\caption{Stellar mass errors versus stellar mass in the mock sample (for clarity, only a third of the data points are shown), together with the mean relation in bins of stellar mass (red points and errorbars).}
\label{fig:ems_ms}
\end{figure}

The \citep{Behroozi13} relations are theoretical, unbiased mean values of $M_*$, given a specific halo mass $M_h$. In any real survey, this relationship will be biased by intrinsic scatter and observational errors (B13,\citet{Leauthaud2012}). 
To quantify this bias for the COSMOS catalogue, we generated a random Monte-Carlo sample of 12,000 halos selected halo mass function given in \citet{Tinker08}. ``True'' stellar masses were calculated for these objects using the SHMR given above. We then added intrinsic scatter and random errors to each stellar mass 50 times independently, to simulate an ``observed'' stellar mass sample. The intrinsic scatter in the SHMR is about 0.14--0.2 dex at redshift of 0 \citep{More2009, Yang2009, Reddick2013}, and there is no evidence for any trend with mass, at least down to halo masses of $10^{12}$ $M_\odot$ \citep{Reddick2013, Behroozi13}. Thus we added an intrinsic scatter of 0.15 to all stellar masses derived for our ``observed'' sample. Average observational errors, as a function of stellar mass, were estimated directly from the COSMOS 2015 catalogue, as shown in Fig.~\ref{fig:ems_ms}. We added these in quadrature to determine the final stellar masses of the mock sample. 

\begin{figure}
\centering
\includegraphics[width=\columnwidth]{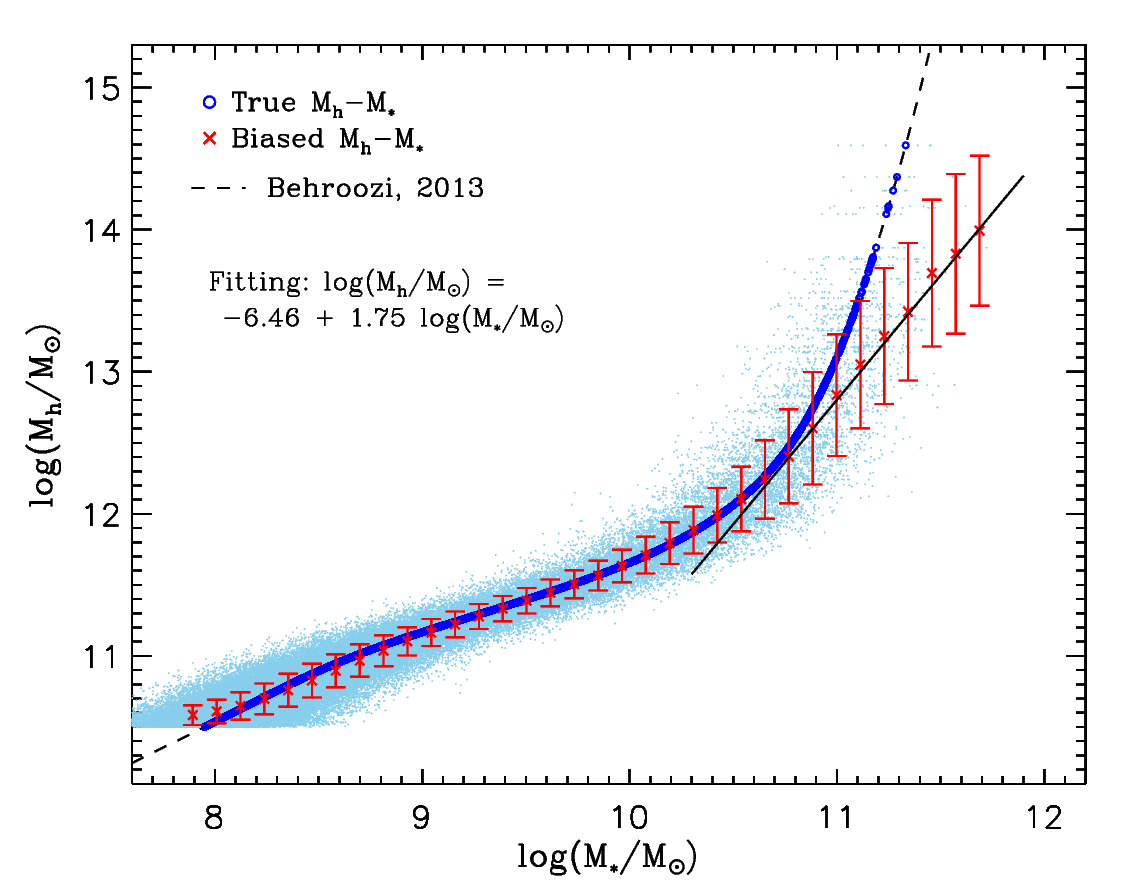}
\caption{Halo-to-stellar mass relation of the Monte-Carlo samples. The dark blue dots are our Monte-Carlo halo mass sample with stellar masses assigned using B13. The light blue points show the ``observed'' stellar masses of the sample after adding intrinsic scatter and observational errors. (Note that only 15\%\ of the mock sample is shown for clarity.) The red dots with error bars are the average halo mass in each bin of ``observed'' stellar mass. The solid black line is a linear fit to the red dots (in $\log(M_*/M_\cdot)$), over the range 10.7--11.8.}
\label{fig:SHMR_bias}
\end{figure}

The light blue points in Fig.~\ref{fig:SHMR_bias} show the ``observed'' stellar masses of the mock sample, after adding the intrinsic scatter and observational errors. The dark blue points show the underlying ``true'' stellar masses, while the black dashed line shows the theoretical SHMR from B13. The red points show the mean halo mass in each ``observed'' stellar mass bin. We can see that the ``observed'' SHMR follows the theoretical SHMR reasonably well at low masses, but departs from it at the high masses. Given this pattern, we will use a two-part SHMR to assign halo masses in our work. For stellar masses of $10^{10.7} M_\odot$ or less, we use an unmodified B13 SHMR, while for masses greater than $10^{10.7}$, we use the linear fit to the average ``observed'' values listed and shown as a black solid line on the plot.

\section{Measuring the Masking and field boundaries}
\label{apd:masking}
Regions of the COSMOS field have poor photometry in one or more bands, due to contamination from bright stars, internal reflections, or other artifacts. Data from these regions are tagged with a ``masking'' flag (``{\tt FLAG\_PETER}'' in the COSMOS 2015 catalogue), which can be used to exclude those data from further analysis. The shape of these masked regions, together with the field boundaries, needs to be measured to determine the area completeness $\eta$ around any given primary. Although detailed mask files are available for the COSMOS field, we found it less computationally demanding for our work to use a single, approximate mask image with coarser spatial sampling. We use the method described in \citet{Xi18} to generate this global mask. First, a coarse map consisting of 390 $\times$ 390 cells is constructed, covering the whole COSMOS field. We search for objects in each cell of this map, to determine whether it should be included or masked out. In the first round, any cell with one or fewer objects counts is selected as a potential masked region. In a second round, these candidates are confirmed as masked if they have one or more neighbouring cells with no counts. This two-step selection process reduces to 0.0026\%\  the probability of artificially eliminating cells due to Poisson fluctuations in their object counts. The map resolution and count threshold were determined empirically after testing various resolutions from 200 $\times$ 200 to 600 $\times$ 600, with different thresholds in each case. We found that the effect on the clustering signal of variations in the masking parameters is small, producing variations in $S_{\rm halo}$ of roughly 5\% or less. The final resolution was selected to provide the most accurate overall mask, relative to the full images.

Given a single global mask for the COSMOS field, we then generated a large, random sample of points, and used the distribution of the points around each primary to estimate its area completeness as a function of projected separation. For each galaxy, we counted the number of random samples in projected radial bins with and without applying the masks and boundaries. Each bin had a size of 0.2 $R_{\rm vir}$ of the galaxy, up to 3.6 $R_{\rm vir}$. The area completeness is then
\begin{equation}
\eta(R_P) \equiv \frac{A_{\rm M}(R_P)}{A_{\rm T}(R_P)}=\frac{N_{\rm M}(R_P)}{N_{\rm T}(R_P)}
\end{equation}
where $A_{\rm M}$ and $A_{T}$ are the masked and total areas, and $N_{\rm M}$ and $N_{\rm T}$ are the random counts with and without masking.

Besides the area completeness in individual radial bins, we also measured the total area completeness of each primary within 3.0 $R_{\rm vir}$. Galaxies with poor completeness were excluded from the primary sample, as described in the main text.

\section{Comparing Background Estimation Methods A and B}
\label{apd:method AvsB}
 While Method A is simple to implement, it may introduce systematic biases in the inferred satellite properties, as described in section \ref{subsec:methodA}. Method B removes the contribution from the background statistically, and should produce less biased, albeit noisier, results.

\begin{figure}
\centering
\includegraphics[width=\columnwidth]{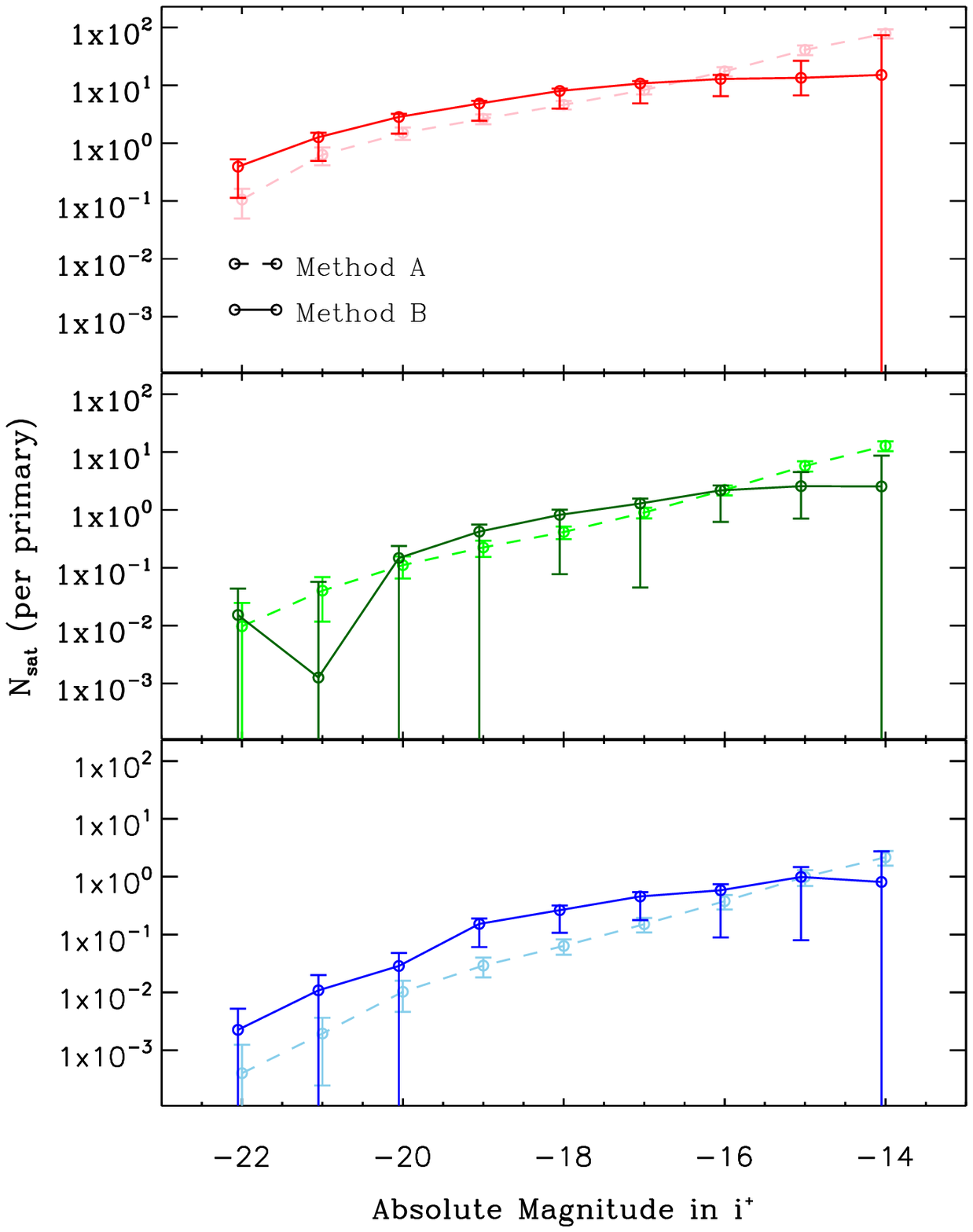}
\caption{Cumulative satellite luminosity functions estimated using methods A (dashed) and B (solid), for primaries in three halo mass bins (as in Fig.~\ref{fig:abmf_3mh_B}). 
}
\label{fig:abmf_3mh_A}
\end{figure}

Fig.~\ref{fig:abmf_3mh_A} shows the (cumulative) satellite luminosity function for three sets of primaries with different halo mass ranges, using methods A (dashed lines) and B (solid lines). Overall, method A produces a luminosity function with a steeper slope, that continues to rise at faint magnitudes, whereas for method B, the cumulative luminosity function flattens. As shown in Fig.~\ref{fig:abmf_3mh_B}, the method B results are in better agreement with Local Group data. 

%%%%%%%%%%%%%%%%%%%%%%%%%%%%%%%%%%%%%%%%%%%%%%%%%%

% Don't change these lines
\bsp% typesetting comment
\label{lastpage}
\end{document}